\documentclass[preprint,amsfonts,amssymb,amsmath,eqsecnum,%
tightenlines,nofootinbib,showpacs,showkeys]{revtex4}

\usepackage{amsthm}

\newcommand{\cH}{\mathcal{H}}
\newcommand{\cM}{\mathcal{M}}
\newcommand{\cN}{\mathcal{N}}

\newcommand{\cV}{\mathcal{V}}

\newcommand{\bH}{\mathbf{H}}
\newcommand{\bM}{\mathbf{M}}
\newcommand{\bS}{\mathbf{S}}
\newcommand{\tH}{\tilde{H}}
\newcommand{\tI}{\tilde{I}}
\newcommand{\tP}{\tilde{P}}
\newcommand{\tcV}{\tilde{\cV}}

\newcommand{\bbR}{\mathbb{R}}
\newcommand{\bbC}{\mathbb{C}}
\newcommand{\Lsl}{\mathfrak{sl}}
\newcommand{\iD}{\mathit{\Delta}}
\newcommand{\Int}{\int\!\!}

\newcommand{\equal}{\phantom{=\ }}

\begin{document}

\title{Type A $\cN$-fold Supersymmetry and Generalized Bender--Dunne
Polynomials}
\author{Toshiaki Tanaka}
\email{totanaka@yukawa.kyoto-u.ac.jp}
\affiliation{Department of Physics,
Graduate School of Science,\\
Osaka University, Toyonaka,
Osaka 560-0043, Japan}
\altaffiliation{Present address: Yukawa Institute for
Theoretical Physics, Kyoto University,
Kyoto 606-8502, Japan}


\begin{abstract}
We derive the necessary and sufficient condition for type A $\cN$-fold
supersymmetry by direct calculation of the intertwining relation and
show the complete equivalence between this analytic construction and
the $\Lsl(2)$ construction based on quasi-solvability.
An intimate relation between the pair of algebraic Hamiltonians is found.
The classification problem on type A $\cN$-fold supersymmetric
models is investigated by considering the invariance of both
the Hamiltonians and $\cN$-fold supercharge under the $GL(2,K)$
transformation. We generalize the Bender--Dunne polynomials to all the
type A $\cN$-fold supersymmetric models without requiring the
normalizability of the solvable sector. Although there is a case where
weak orthogonality of them is not guaranteed, this fact does not cause
any difficulty on the generalization. It is shown that the
anti-commutator of the type A $\cN$-fold supercharges is expressed as
the critical polynomial of them in the original Hamiltonian, from which
we establish the complete type A $\cN$-fold superalgebra. A novel
interpretation of the critical polynomials in view of polynomial
invariants is given.
\end{abstract}

\pacs{02.10.De; 03.65.Fd; 11.30.Na; 11.30.Pb}
\keywords{quantum mechanics; quasi-solvability; $\cN$-fold supersymmetry;
intertwining relation; Bender--Dunne polynomials;
polynomial invariants}

\preprint{OU-HEP 426}

\maketitle

\section{\label{sec:intro}Introduction}

The concept of symmetry has played a central role in modern theoretical
physics. A discovery of a new symmetry enlarges our ability and
possibility to describe new phenomena both in the physical nature
and in mathematical models. Conversely, it may be almost certain that
there is an underlying symmetry if a system under consideration exhibits
a significant property that is not shared in the generic cases.
Actually, $\cN$-fold supersymmetry was discovered from the observation
of the disappearance of the leading divergence of the perturbation
series for the specific energy levels of a quantum mechanical model
at the particular values of a parameter involved in the
model~\cite{AKOSW2}. This symmetry is a generalization of the ordinary
supersymmetry in one-dimensional quantum mechanics~\cite{Witte1,Witte2}
and is characterized by $\cN$th order derivative supercharges.
Similar generalizations were found and investigated in various different
contexts~\cite{AnIoSp1,AnIoCaDe1,AnIoNi1,AnIoNi2,BaSa1,Samso1,BaSa2,%
Samso2,Ferna1,Ortiz1,BaGaBhMi1,FeHu1,FeNeNi1,Plyus1,KlPl1}.
A generalization to $\cN$th order derivative supercharges for
all $\cN$ is, however, difficult and most of the investigations by
the other authors were limited to the cases of the second-order
generalization and to the cases constructed by the factorization
method. Recently, we found another $\cN$-fold supersymmetric
model~\cite{ASTY} and further found general
forms of an $\cN$-fold supersymmetric family which we have called
type A~\cite{AST1}. On the other hand, several quasi-solvable
models~\cite{TuUs1,Ushve} were found to have $\cN$-fold supersymmetry
in Refs.~\cite{KlPl2,KlPl3,DoDuTa1,DoDuTa2} in addition to those
reported in Refs.~\cite{AKOSW2,ASTY}. Then, it was proved in generic
way~\cite{AST2} that $\cN$-fold supersymmetry is essentially equivalent
to quasi-solvability. Furthermore, it was also proved~\cite{ANST1}
that the type A $\cN$-fold supersymmetric models are essentially
equivalent to the quasi-solvable models constructed by $\Lsl(2)$
generators~\cite{Turbi1}. Dynamical properties of the $\cN$-fold
supersymmetric models were discussed in Ref.~\cite{AST2} and investigated
for a couple of models in Ref.~\cite{ST1}. The nonperturbative analyses
carried out in Ref.~\cite{ST1} together with those in Ref.~\cite{AKOSW2}
revealed several significant properties that the type A $\cN$-fold
supersymmetric models share. Especially, we clarified the important role
of the normalizability of the solvable sector, which is crucial for
the dynamical $\cN$-fold supersymmetry breaking but has rarely discussed
by the other authors. Furthermore, it was argued that $\cN$-fold
supersymmetry is not only sufficient but may also be necessary for the
existence of convergent perturbation series.
Up to now, most of the quasi-solvable
one-dimensional quantum mechanical systems belong to the $\Lsl(2)$
quasi-solvable models. Several new findings have been reported in
connection with them. One of the examples is the new orthogonal
polynomials firstly found by Bender and Dunne~\cite{BeDu1}. The idea
was soon generalized to all the $\Lsl(2)$ quasi-exactly solvable
models~\cite{KrUsWa1,FiLoRo1}. Furthermore, several realistic physical
systems have been found, which can be reduced to one-dimensional
quasi-solvable models~\cite{KlPl3,SaGh1,Taut1,Taut2,Turbi3,Taut3,ViPi1,%
Taut4,HoKh1,ChHo1,ChHo2}.

In this article, we will report the general aspects of type A $\cN$-fold
supersymmetry, most of which remain unsolved yet. In Ref.~\cite{ANST1},
it was discussed that, in view of the $\Lsl(2)$ construction of type
A $\cN$-fold supersymmetry, the condition for type A $\cN$-fold
supersymmetry derived in Ref.~\cite{AST1} provides only a sufficient
but not necessary conditions. On the other hand, only one of the pair
of the type A Hamiltonians was investigated in Ref.~\cite{ANST1} and
relations between the pair in view of the $\Lsl(2)$ structure have not
been known. In other words, the equivalence between the analytic
construction in Ref.~\cite{AST1} and the algebraic one in Ref.~\cite{ANST1}
has not been established. Another problem to be solved is the
classification problem. In Ref.~\cite{ANST2}, we attempted to classify
the type A $\cN$-fold supersymmetric models by considering the invariance
under the linear transformations but found to be incomplete. Later
we found that the complete classification of the $\Lsl(2)$ quasi-solvable
models was already achieved in Ref.~\cite{LoKaOl3,LoKaOl4} by
the consideration of more extensive $GL(2,\bbR)$ transformations.
Therefore, we have recognized that the complete classification of the
type A models should be done by investigating the invariance of both
the Hamiltonians and $\cN$-fold supercharge under the $GL(2,\bbR)$
transformations. Lastly, the structure of the anti-commutator of the
type A $\cN$-fold supercharges, which we have called mother Hamiltonian,
has not been investigated yet at all and thus we have not established
the complete type A $\cN$-fold superalgebra. We will give the
complete answers to the above problems in this article.

The article is organized as follows. In the next section, we give
a brief review on the definition and the general
properties of $\cN$-fold supersymmetry needed in this article.
In Section~\ref{sec:typea}, we define type A $\cN$-fold supersymmetry
and give two different approaches, namely, analytic and algebraic
approaches to construct the type A $\cN$-fold supersymmetric models.
The classification problem is discussed in Section~\ref{sec:class}.
By considering the invariance property of both the Hamiltonians and
$\cN$-fold supercharge, we obtain the complete answer to the problem. In
Section~\ref{sec:2f2dp}, we investigate 2-fold supersymmetry with
an emphasis on the uniqueness of supercharges and the polynomiality
of the 2-fold superalgebras. A novel feature of weak quasi-solvability
defined in Section~\ref{sec:nsusy} is discussed.
In Section~\ref{sec:mhamBD}, we first show
the polynomiality of the type A $\cN$-fold superalgebras. Then, by
a suitable generalization of the original idea of the Bender--Dunne
polynomials to all the type A $\cN$-fold supersymmetric models, we
show that the anti-commutators of the type A $\cN$-fold supercharges
are expressed as the critical polynomials of the generalized Bender--Dunne
polynomials. We claim that both the normalizability of the solvable
sector and the weak orthogonality of the polynomials do not play
essential roles on the generalization. A novel interpretation of the
critical polynomials is given in view of polynomial invariants.
Interesting future problems are discussed in the final
section. In Appendix~\ref{app:invsp}, we summarize the results of
the invariant theory of polynomial systems needed in this article.

\section{\label{sec:nsusy}$\cN$-fold Supersymmetry}

\subsection{\label{ssec:defnss}Definition}

We start with defining $\cN$-fold supersymmetry in one dimensional
quantum mechanics. Let us introduce a bosonic coordinate $q$ and
fermionic coordinates $\psi$ and $\psi^{\dagger}$ satisfying,
\begin{align}
\{\psi,\psi\}=\{\psi^{\dagger},\psi^{\dagger}\}=0,
\qquad \{\psi,\psi^{\dagger}\}=1.
\label{eqn:fermi}
\end{align}
Hamiltonian $\bH_{\cN}$ is given by,
\begin{align}
\bH_{\cN}=H_{\cN}^{-}\,\psi\psi^{\dagger}+H_{\cN}^{+}\,\psi^{\dagger}\psi,
\label{eqn:nfham}
\end{align}
where $H_{\cN}^{\pm}$ are ordinary scalar Hamiltonians:
\begin{align}
H_{\cN}^{\pm}=\frac{1}{2}p^{2}+V_{\cN}^{\pm}(q),
\label{eqn:ordiH}
\end{align}
with $p=-id/dq$. $\cN$-fold supercharges $Q_{\cN}$ and $Q_{\cN}^{\dagger}$
are introduced by,
\begin{align}
Q_{\cN}=P_{\cN}^{\dagger}\,\psi,\qquad
 Q_{\cN}^{\dagger}=P_{\cN}\,\psi^{\dagger},
\end{align}
where $P_{\cN}$ is given by a polynomial of $\cN$th degree in $p$:
\begin{align}
P_{\cN}=p^{\cN}+w_{\cN-1}(q)p^{\cN-1}+\dots+w_{1}(q)p+w_{0}(q),
\label{eqn:nfsch}
\end{align}
that is, $P_{\cN}$ is an $\cN$th-order linear differential operator.
Then, the system (\ref{eqn:nfham}) is defined to be \textit{$\cN$-fold
supersymmetric} if the following algebra holds:
\begin{gather}
\{Q_{\cN},Q_{\cN}\}=\{Q_{\cN}^{\dagger},Q_{\cN}^{\dagger}\}=0,\\
[Q_{\cN},\bH_{\cN}]=[Q_{\cN}^{\dagger},\bH_{\cN}]=0.
\label{eqn:nfalg2}
\end{gather}
The former relation is trivial due to Eq.~(\ref{eqn:fermi}) while the
latter is equivalent to the following intertwining relations:
\begin{align}
P_{\cN}H_{\cN}^{-}-H_{\cN}^{+}P_{\cN}=0,\qquad
P_{\cN}^{\dagger}H_{\cN}^{+}-H_{\cN}^{-}P_{\cN}^{\dagger}=0.
\label{eqn:inter}
\end{align}
Therefore, the relation (\ref{eqn:inter}) gives the condition for the
system $\bH_{\cN}$ to be $\cN$-fold supersymmetric. From the definition,
it is evident that $\cN$-fold supersymmetry reduces to the ordinary
supersymmetry \cite{Witte1,Witte2} when $\cN=1$.

\subsection{\label{ssec:qsolve}Quasi-solvability}

The $\cN$-fold supersymmetric models defined above have several
significant properties similar to those of the ordinary supersymmetric
models \cite{AST2}. One of the most notable ones is
\text{quasi-solvability} \cite{TuUs1,Ushve}. A linear differential
operator $H$ of a single variable $q$ is said to be \textit{quasi-solvable}
if it preserves a finite dimensional functional space $\cV_{\cN}$ whose
basis admits an explicit analytic form:
\begin{align}
H\cV_{\cN}&\subset\cV_{\cN}, & \dim\cV_{\cN}&=n(\cN)<\infty,
 & \cV_{\cN}&=\text{span}\,\left\{\phi_{1}(q),\dots,\phi_{n(\cN)}(q)
 \right\}.
\label{eqn:defqs}
\end{align}
An immediate consequence of the above definition of quasi-solvability
is that, since we can calculate finite dimensional matrix elements
$\bS_{k,l}$ defined by,
\begin{align}
H\phi_{k}=\sum_{l=1}^{n(\cN)}\bS_{k,l}\phi_{l}\qquad
 \bigl(k=1,\dots,n(\cN)\bigr),
\end{align}
we can diagonalize the operator $H$ and obtain the spectra of it in the
space $\cV_{\cN}$, at least, algebraically. Furthermore, if the space
$\cV_{\cN}$ is a subspace of a Hilbert space $L^{2}(S)$ ($S\subset\bbR$)
on which the operator $H$ is naturally defined, the solvable spectra
and the corresponding vectors of $\cV_{\cN}$ give the \textit{exact}
eigenvalues and eigenfunctions of $H$, respectively. In this case, the
operator $H$ is said to be \textit{quasi-exactly solvable}. The role of
the normalizability of the solvable sector is investigated in view of
dynamical properties in Ref.~\cite{AKOSW2,ST1}.
To construct a quasi-solvable model, it is convenient to introduce
an $\cN$th-order linear differential
operator $P$ and define the vector space $\cV_{\cN}$ as,
\begin{align}
\cV_{\cN}=\ker P.
\label{eqn:svspc}
\end{align}
Now, it is easy to see that an operator $H$ is quasi-solvable with the
solvable sector (\ref{eqn:svspc}) if the following quasi-solvability
condition holds \cite{AST1}:
\begin{align}
PH\cV_{\cN}=0.
\label{eqn:qscon}
\end{align}
This formulation enables us to clarify the relation between
quasi-solvability and $\cN$-fold supersymmetry. Indeed, it can be easily
shown that all the quasi-solvable models satisfying Eq.~(\ref{eqn:qscon})
are $\cN$-fold supersymmetric if we set $P_{\cN}=P$, $H_{\cN}^{-}=H$
and $H_{\cN}^{+}=H+iw'_{\cN-1}$, where $w_{\cN-1}$ is defined in
Eq.~(\ref{eqn:nfsch}). The converse is also true.
From the intertwining relations (\ref{eqn:inter}), we find that all the
$\cN$-fold supersymmetric systems are quasi-solvable: the quasi-solvability
condition (\ref{eqn:qscon}) holds for $H=H_{\cN}^{-}(H_{\cN}^{+})$ and
$P=P_{\cN}(P_{\cN}^{\dagger})$, respectively. Therefore, if we define,
\begin{subequations}
\begin{align}
\cV_{\cN}^{-}&=\ker P_{\cN}=\text{span}\,\{\phi_{n}^{-}:n=1,\dots,\cN\},
\label{eqn:dfVn-}\\
\cV_{\cN}^{+}&=\ker P_{\cN}^{\dagger}=\text{span}\,\{\phi_{n}^{+}:
 n=1,\dots,\cN\},
\end{align}
\end{subequations}
we have $H_{\cN}^{\pm}\cV_{\cN}^{\pm}\subset\cV_{\cN}^{\pm}$ and
the following Schr\"{o}dinger equations on the subspaces
$\cV_{\cN}^{\pm}$:
\begin{align}
H_{\cN}^{\pm}\phi_{k}^{\pm}=\sum_{l=1}^{\cN}\bS_{k,l}^{\pm}\phi_{l}^{\pm}
=E_{k}^{\pm}\phi_{k}^{\pm}\qquad (k=1,\dots,\cN).
\label{eqn:matS1}
\end{align}
The spectra $E^{\pm}$ are determined from the characteristic equations
for the matrices $\bS^{\pm}$:
\begin{align}
\det\bM_{\cN}^{\pm}(E^{\pm})=0,\qquad\bM_{\cN}^{\pm}(\lambda)
 =2(\lambda\mathbf{I}-\bS^{\pm}).
\label{eqn:chaeq}
\end{align}
We should note that, for a given operator $P$, we cannot always
obtain analytic solutions of Eq.~(\ref{eqn:svspc}). Therefore,
quasi-solvability formulated from Eqs.~(\ref{eqn:svspc}) and
(\ref{eqn:qscon}) is less restrictive than the one defined by
Eq.~(\ref{eqn:defqs}). In the situation where this difference is
crucial, we may be better to call the less restrictive case
\textit{weakly} quasi-solvable. With this terminology, we say
more correctly that $\cN$-fold supersymmetry is equivalent to weak
quasi-solvability. In Section~\ref{sec:2f2dp}, we will discuss
weak quasi-solvability again.

\subsection{\label{ssec:motham}Mother Hamiltonian}

In the ordinary supersymmetry, the anti-commutator of the supercharges
corresponds to the Hamiltonian. However, it is not the case in $\cN$-fold
supersymmetry. This is because $\{ Q_{\cN}^{\dagger}, Q_{\cN}\}$ is
now a $2\cN$th-order differential operator. The half of the
anti-commutator is called \textit{mother Hamiltonian} and is denoted
by $\cH_{\cN}$:
\begin{align}
\cH_{\cN}=\frac{1}{2}\{Q_{\cN}^{\dagger},Q_{\cN}\}.
\label{eqn:moham}
\end{align}
An immediate consequence of the above definition is that the mother
Hamiltonian always commutes with the $\cN$-fold supercharges, that is,
it is $\cN$-fold supersymmetric:
\begin{align}
[Q_{\cN},\cH_{\cN}]=[Q_{\cN}^{\dagger},\cH_{\cN}]=0.
\label{eqn:mhrel1}
\end{align}
Furthermore, if the original Hamiltonian $\bH_{\cN}$ is $\cN$-fold
supersymmetric, the mother Hamiltonian also commutes with $\bH_{\cN}$
due to the relation (\ref{eqn:nfalg2}):
\begin{align}
[\cH_{\cN},\bH_{\cN}]=0.
\end{align}
From the above relations, it is expected that the mother Hamiltonian
$\cH_{\cN}$ has an intimate relation with the original Hamiltonian
$\bH_{\cN}$. Indeed, it was shown \cite{AST2} that, if the $\cN$-fold
supercharges $Q_{\cN}$ and $Q_{\cN}^{\dagger}$ are uniquely determined
and $H_{\cN}^{-}\neq H_{\cN}^{+}$,
$\cH_{\cN}$ is expressed as the characteristic polynomial of $\cN$-th
degree for $\bS^{\pm}$ appeared in Eq.~(\ref{eqn:chaeq}) with the
argument replaced by $\bH_{\cN}$:
\begin{align}
\cH_{\cN}=\frac{1}{2}\det\bM_{\cN}^{+}(\bH_{\cN})
 =\frac{1}{2}\det\bM_{\cN}^{-}(\bH_{\cN}).
\label{eqn:mhplh}
\end{align}

\section{\label{sec:typea}Type A $\cN$-fold Supersymmetric Models}

In contrast to the ordinary supersymmetric quantum mechanics, the
construction of an $\cN$-fold supersymmetric model is a non-trivial
problem. In the case of ordinary supersymmetry, the mother Hamiltonian
(\ref{eqn:moham}) defined by the anti-commutator of the supercharges
is a desirable Schr\"{o}dinger operator of the form (\ref{eqn:nfham})
and thus can be identified as a supersymmetric Hamiltonian due to the
relation (\ref{eqn:mhrel1}). In the case of 2-fold supersymmetry,
the condition (\ref{eqn:inter}) can be completely solved and thus the
most general form of the 2-fold supersymmetric Hamiltonian and the 2-fold
supercharge have been known~\cite{AnIoCaDe1,AnIoNi1,AST2}. However,
we can hardly solve the condition (\ref{eqn:inter}) for $\cN\ge 3$ in
general. Nevertheless, we have found a special case where the condition
(\ref{eqn:inter}) can be solved for arbitrary $\cN$. We have called
this case \textit{type A}~\cite{AST1}.
The type A $\cN$-fold supercharge is defined as the following special
form of $\cN$th-order linear differential operator:
\begin{align}
P_{\cN}&=\prod_{k=-(\cN-1)/2}^{(\cN-1)/2}
 \bigl(p-iW(q)+ikE(q)\bigr)\notag\\
&=\left(p-iW(q)+i\frac{\cN-1}{2}E(q)\right)
 \left(p-iW(q)+i\frac{\cN-3}{2}E(q)\right)\times\cdots\notag\\
&\equal \dots\times\left(p-iW(q)-i\frac{\cN-3}{2}E(q)\right)
 \left(p-iW(q)-i\frac{\cN-1}{2}E(q)\right).
\label{eqn:typea}
\end{align}
In the above definition, we note that $E(q)$ and $W(q)$ correspond
to $\widetilde{E}(q)$ and $\widetilde{W}(q)$, respectively, in the
previous articles~\cite{AST1,AST2,ANST1,ANST2}. We will use tildes
for another particular purpose (see Eqs.~(\ref{eqns:gtPaH}) and below)
and thus we do not follow the old notation anymore.
A system (\ref{eqn:nfham}) is said to be \textit{type A $\cN$-fold
supersymmetric}
if the condition (\ref{eqn:inter}) is fulfilled with this type A
$\cN$-fold supercharge. There are two ways to construct a type A model,
namely, \textit{analytic} and \textit{algebraic} constructions. The former
is to solve the intertwining relation (\ref{eqn:inter}) directly while
the latter is to solve the quasi-solvability condition (\ref{eqn:qscon}).
In the following sections, we will first show the analytic construction
and next the algebraic one.

\subsection{\label{ssec:analyt}Analytic Construction}

The condition for type A $\cN$-fold supersymmetry was firstly investigated
with the aid of induction in Ref.~\cite{AST1}. Later, it was reexamined
in the context of the algebraic construction \cite{ANST1}. Then, it has
turned out that the set of the conditions derived in Ref.~\cite{AST1}
only gives a sufficient one. The origin of the defect was explained in
Ref.~\cite{ANST1}. In the following, we will give an improved
direct proof of the necessary and sufficient condition for type A
$\cN$-fold supersymmetry.

The necessary and sufficient condition for the system (\ref{eqn:nfham})
to be $\cN$-fold supersymmetry with respect to the type A $\cN$-fold
supercharge (\ref{eqn:typea}) is the following:
\begin{subequations}
\label{eqns:acond}
\begin{gather}
V_{\cN}^{\pm}(q)=\frac{1}{2}W(q)^{2}+\frac{1}{2}v_{\cN}^{\pm}(q),
\quad v_{\cN}^{\pm}(q)=\frac{\cN^{2}-1}{12}\left(E(q)^{2}-2E'(q)\right)
\pm\cN W'(q)-2R,
\label{eqn:acond1}
\\
\left(\frac{d}{dq}-E(q)\right)\frac{d}{dq}\left(\frac{d}{dq}+E(q)\right)
 W(q)=0\qquad\text{for}\qquad\cN\ge 2,
\label{eqn:acond2}
\\
\left(\frac{d}{dq}-2E(q)\right)\left(\frac{d}{dq}-E(q)\right)\frac{d}{dq}
\left(\frac{d}{dq}+E(q)\right)E(q)=0
\qquad\text{for}\qquad\cN\ge 3,
\label{eqn:acond3}
\end{gather}
\end{subequations}
where $R$ is an arbitrary constant.

\begin{proof}

The proposition will be proved by induction. At first, we make
\textit{gauge} transformations on $P_{\cN}$ and $H_{\cN}^{\pm}$
to facilitate the calculations, as follows:
\begin{subequations}
\label{eqns:gtPaH}
\begin{gather}
\tP_{\cN}=i^{\cN}(G_{\cN}U)P_{\cN}(G_{\cN}U)^{-1}
 =\prod_{k=0}^{\cN-1}(\partial-kE),
\\
\tH_{\cN}^{\pm}=(G_{\cN}U)H_{\cN}^{\pm}(G_{\cN}U)^{-1},
\end{gather}
\end{subequations}
where $G_{\cN}$ and $U$ are defined by,
\begin{align}
G_{\cN}=\exp\left(\frac{\cN-1}{2}\Int dq E(q)\right),\qquad
 U=\exp\left(\Int dq W(q)\right).
\end{align}
In the above and hereafter, we attach tildes to operators, vectors and
vector spaces to indicate that they are quantities gauge-transformed
with the gauge factor $G_{\cN}U$. We set,
\begin{align}
\tI_{\cN}=2(\tP_{\cN}\tH_{\cN}^{-}-\tH_{\cN}^{+}\tP_{\cN}).
\end{align}
We assume that $\tI_{\cN}=0$ for a natural number $\cN$ if the set
of the conditions (\ref{eqns:acond}) holds for this $\cN$. Then,
we will prove that $\tI_{\cN+1}=0$ if and only if the set of the
conditions (\ref{eqns:acond}) holds for $\cN$ replaced by $\cN+1$.
From the following relation:
\begin{align}
2(G_{\cN+1}U)H_{\cN}^{\pm}(G_{\cN+1}U)^{-1}=2\tH_{\cN}^{\pm}
 +E\partial+\frac{1}{2}E'-\frac{2\cN-1}{4}E^{2}-EW,
\end{align}
we obtain,
\begin{align}
2\tH_{\cN+1}^{\pm}&=(G_{\cN+1}U)(2H_{\cN}^{\pm}-v_{\cN}^{\pm}
 +v_{\cN+1}^{\pm})(G_{\cN+1}U)^{-1}\notag\\
&=2\tH_{\cN}^{\pm}+E\partial +\frac{1}{2}E'-\frac{2\cN-1}{4}E^{2}
 -EW-v_{\cN}^{\pm}+v_{\cN+1}^{\pm}\notag\\
&=2\tH_{\cN}^{\pm}+E\partial +u_{\cN+1}^{\pm}-u_{\cN}^{\pm},
\label{eqn:tHn+1}
\end{align}
where $u_{\cN}^{\pm}$ are defined by,
\begin{align}
u_{\cN}^{\pm}=v_{\cN}^{\pm}+W'+\frac{\cN-1}{2}E'
 -\frac{(\cN-1)^{2}}{4}E^{2}-(\cN-1)EW.
\label{eqn:defuN}
\end{align}
In the above, we note that $v_{\cN}^{\pm}$ are given by
Eq.~(\ref{eqn:acond1}) from the inductive assumption while
$v_{\cN+1}^{\pm}$ are unknown functions to be determined.
From Eq.~(\ref{eqn:tHn+1}), we obtain,
\begin{align}
\tI_{\cN+1}&=\tP_{\cN+1}\left(2\tH_{\cN}^{-}+E\partial +u_{\cN+1}^{-}
 -u_{\cN}^{-}\right)-\left(2\tH_{\cN}^{+}+E\partial +u_{\cN+1}^{+}
 -u_{\cN}^{+}\right)\tP_{\cN+1}\notag\\
&=\left(u_{\cN+1}^{-}-u_{\cN}^{-}-u_{\cN+1}^{+}+u_{\cN}^{+}\right)
 \tP_{\cN+1}-2 [\tH_{\cN}^{+}, \partial-\cN E ]\tP_{\cN}\notag\\
&\equal +\tP_{\cN+1}E\partial -E\partial\tP_{\cN+1}+[
 \tP_{\cN+1}, u_{\cN+1}^{-}-u_{\cN}^{-}].
\label{eqn:In+11}
\end{align}
The last four terms in the r.h.s. of Eq.~(\ref{eqn:In+11}) are
calculated as follows:
\begin{subequations}
\label{eqns:form4}
\begin{align}
2 [\tH_{\cN}^{+},\partial-\cN E ]
&=\bigl( (\cN+1)E'-2W'\bigr) (\partial-\cN E)\notag\\
&\equal -\left(u_{\cN}^{+}+2\cN WE-\cN E^{2}-\cN E'\right)',\\
\tP_{\cN+1}E\partial
&=E\tP_{\cN+2}+\sum_{n=1}^{\cN+1}\prod_{k=n}^{\cN}(\partial-kE)
 E_{(-1)}\tP_{n},\notag\\
&=E\tP_{\cN+2}+(\cN+1)E_{(-1)}\tP_{\cN+1}+\sum_{n=1}^{\cN}n
 \prod_{k=n+1}^{\cN}(\partial-kE)E_{(0,-1)}\tP_{n},\\
E\partial\tP_{\cN+1}
&=E\tP_{\cN+2}+(\cN+1)E^{2}\tP_{\cN+1},\\
[\tP_{\cN+1}, u_{\cN+1}^{-}-u_{\cN}^{-}]
&=\sum_{n=0}^{\cN}\prod_{k=n+1}^{\cN}(\partial-kE)
 (u_{\cN+1}^{-}-u_{\cN}^{-})_{(0)}\tP_{n},
\end{align}
\end{subequations}
In the above and hereafter, we employ the following abbreviations:
\begin{align}
f_{(k)}=(\partial-kE)f,\qquad f_{(k,\dots)}=(\partial-kE)f_{(\ldots)}.
\label{eqn:abbre}
\end{align}
The following formula is useful for the calculations:
\begin{align}
\prod_{k=n+l}^{\cN}(\partial-kE)f=f\prod_{k=n}^{\cN-l}(\partial-kE)
 +\sum_{n'=n}^{\cN-1}\prod_{k=n'+l+1}^{\cN}(\partial-kE)f_{(l)}
 \prod_{k'=n}^{n'-1}(\partial-k'E).
\label{eqn:formp}
\end{align}
Substituting Eqs.~(\ref{eqns:form4}) for Eq.~(\ref{eqn:In+11}), we have,
\begin{align}
\tI_{\cN+1}&=\left(2W'+u_{\cN+1}^{-}-u_{\cN}^{-}-u_{\cN+1}^{+}
 +u_{\cN}^{+}\right)\tP_{\cN+1}+\left(u_{\cN}^{+}+2\cN WE-\cN E^{2}
 -\cN E'\right)'\tP_{\cN}\notag\\
&\equal +\sum_{n=0}^{\cN}\prod_{k=n+1}^{\cN}(\partial-kE)
 \left[(u_{\cN+1}^{-}-u_{\cN}^{-})_{(0)}+n E_{(0,-1)}\right]\tP_{n}.
\label{eqn:In+12}
\end{align}
From Eq.~(\ref{eqn:In+12}), we see that $\tI_{\cN+1}$ is at most
($\cN+1$)th-order differential operator. Therefore, $\tI_{\cN+1}=0$ if
and only if all the coefficients of $\partial^{k}$ ($k=0,1,\dots,\cN+1$)
vanish. Only the first term in the r.h.s. of Eq.~(\ref{eqn:In+12})
contains the $\partial^{\cN+1}$ term. Therefore, one of the conditions
for $\tI_{\cN+1}=0$ reads,
\begin{align}
2W'+u_{\cN+1}^{-}-u_{\cN}^{-}-u_{\cN+1}^{+}+u_{\cN}^{+}=0.
\label{eqn:cdI01}
\end{align}
Combining Eq.~(\ref{eqn:cdI01}) with Eqs.~(\ref{eqn:acond1}) and
(\ref{eqn:defuN}), we have,
\begin{align}
v_{\cN+1}^{+}-v_{\cN+1}^{-}=2(\cN+1)W'.
\label{eqn:cdI02}
\end{align}
Applying the formula (\ref{eqn:formp}) and the condition (\ref{eqn:cdI01})
to Eq.~(\ref{eqn:In+12}), we obtain,
\begin{align}
\tI_{\cN+1}&=\left[\frac{\cN(\cN+1)}{2}E_{(0,-1)}+(\cN+1)(u_{\cN+1}^{-}
 -u_{\cN}^{-})_{(0)}+\left(u_{\cN}^{+}+2\cN WE-\cN E^{2}-\cN E'\right)'
 \right]\tP_{\cN}\notag\\
&\equal +\sum_{n=0}^{\cN}\sum_{n'=n}^{\cN-1}\prod_{k=n'+2}^{\cN}
 (\partial-kE)\left[(u_{\cN+1}^{-}-u_{\cN}^{-})_{(1,0)}+n E_{(1,0,-1)}
 \right]\tP_{n'}.
\label{eqn:In+13}
\end{align}
Only the first term in the r.h.s. of Eq.~(\ref{eqn:In+13})
contains the $\partial^{\cN}$ term. Therefore, we obtain another
condition for $\tI_{\cN+1}=0$:
\begin{align}
\frac{\cN(\cN+1)}{2}E_{(0,-1)}+(\cN+1)(u_{\cN+1}^{-}-u_{\cN}^{-})_{(0)}
 +\left(u_{\cN}^{+}+2\cN WE-\cN E^{2}-\cN E'\right)'=0.
\label{eqn:cdI03}
\end{align}
Combining the assumption (\ref{eqn:acond1}) with Eqs.~(\ref{eqn:cdI02})
and (\ref{eqn:cdI03}), we finally have,
\begin{align}
v_{\cN+1}^{\pm}=\frac{(\cN+1)^{2}-1}{12}\left(E^{2}-2E'\right)
 \pm(\cN+1)W'+\text{const.}
\label{eqn:cdI04}
\end{align}
The resulting $v_{\cN+1}^{\pm}$ is nothing but the assumed form
(\ref{eqn:acond1}) with $\cN$ replaced by $\cN+1$. Therefore, the first
condition (\ref{eqn:acond1}) has been proved inductively. Under
the above conditions satisfied, we have from Eqs.~(\ref{eqn:defuN})
and (\ref{eqn:cdI04}),
\begin{align}
u_{\cN+1}^{-}-u_{\cN}^{-}
&=v_{\cN+1}^{-}-v_{\cN}^{-}+\frac{1}{2}E'-\frac{2\cN-1}{4}E^{2}-EW\notag\\
&=-\frac{\cN-1}{3}E_{(-1)}-W_{(-1)}.
\end{align}
Substituting the above for Eq.~(\ref{eqn:In+13}), we obtain,
\begin{align}
\tI_{\cN+1}=-\sum_{n=0}^{\cN}\sum_{n'=n}^{\cN-1}\prod_{k=n'+2}^{\cN}
 (\partial-kE)\left[W_{(1,0,-1)}-\frac{3n-\cN+1}{3}E_{(1,0,-1)}\right]
 \tP_{n'}.
\end{align}
The above expression for $\tI_{\cN+1}$ is further arranged with the aid
of the formula (\ref{eqn:formp}) as follows:
\begin{align}
\tI_{\cN+1}
&=-\sum_{n=0}^{\cN}\sum_{n'=n}^{\cN-1}\left(W_{(1,0,-1)}-\frac{
 3n-\cN+1}{3}E_{(1,0,-1)}\right)\tP_{\cN-1}\notag\\
&\equal -\sum_{n=0}^{\cN}\sum_{n'=n}^{\cN-1}\sum_{n''=n'}^{\cN-2}
 \prod_{k=n''+3}^{\cN}(\partial-kE)\left(W_{(2,1,0,-1)}-\frac{
 3n-\cN+1}{3}E_{(2,1,0,-1)}\right)\tP_{n''}\notag\\
&=-\frac{\cN(\cN+1)}{2}W_{(1,0,-1)}\tP_{\cN-1}\notag\\
&\equal -\frac{1}{2}\sum_{n=0}^{\cN-2}(n+1)(n+2)\prod_{k=n+3}^{\cN}
(\partial-kE)\left(W_{(2,1,0,-1)}-\frac{n-\cN+1}{3}E_{(2,1,0,-1)}\right)
 \tP_{n}.
\end{align}
When $\cN=1$, $\tI_{2}=-W_{(1,0,-1)}$ and thus we obtain,
\begin{align}
\tI_{2}=0\quad\Leftrightarrow\quad
 W_{(1,0,-1)}=\left(\frac{d}{dq}-E\right)\frac{d}{dq}
 \left(\frac{d}{dq}+E\right)W=0,
\end{align}
i.e., the condition (\ref{eqn:acond2}) in addition to
Eq.~(\ref{eqn:acond1}). When $\cN\ge 2$,
the condition (\ref{eqn:acond2}) has been already assumed and thus,
\begin{align}
\tI_{\cN+1}=\frac{1}{6}\sum_{n=0}^{\cN-2}(n+1)(n+2)\prod_{k=n+3}^{\cN}
 (\partial-kE)E_{(2,1,0,-1)}\tP_{n}.
\end{align}
When $\cN=2$, $\tI_{3}=-E_{(2,1,0,-1)}/3$ and thus we obtain,
\begin{align}
\tI_{3}=0\quad\Leftrightarrow\quad
 E_{(2,1,0,-1)}=\left(\frac{d}{dq}-2E\right)\left(\frac{d}{dq}-E\right)
 \frac{d}{dq}\left(\frac{d}{dq}+E\right)E=0,
\end{align}
i.e., the condition (\ref{eqn:acond3}) in addition to
Eqs.~(\ref{eqn:acond1}) and (\ref{eqn:acond2}). When $\cN\ge 3$,
the condition (\ref{eqn:acond3}) has been already assumed too and thus,
\begin{align}
\tI_{\cN+1}=0.
\end{align}
Therefore, no additional condition is required any more.
\end{proof}

From the set of the conditions (\ref{eqns:acond}), we find a procedure
to construct a type A $\cN$-fold supersymmetric model as follows:
\begin{enumerate}
\item Find out a particular solution $E(q)$ of the nonlinear differential
equation (\ref{eqn:acond3}).
\item Substitute the above $E(q)$ for the general solution of
the linear differential equation (\ref{eqn:acond2}) given by,
\begin{align}
W(q)&=C_{1}e^{-\Int dq E(q)}\Int dq\left(e^{\Int dq E(q)}\Int dq\,
 e^{\Int dq E(q)}\right)\notag\\
&\equal +C_{2}\, e^{-\Int dq E(q)}\Int dq\, e^{\Int dq E(q)}
 +C_{3}\, e^{-\Int dq E(q)}\quad (C_{i}=\text{arbitrary constants}),
\end{align}
to obtain $W(q)$.
\item Substitute the above $E(q)$ and $W(q)$ for Eq.~(\ref{eqn:acond1})
to obtain the pair of potentials $V_{\cN}^{\pm}(q)$.
\end{enumerate}

\subsection{\label{ssec:sl2con}$\Lsl(2)$ Construction}

As previously noted, $\cN$-fold supersymmetry is essentially equivalent
to quasi-solvability. Recently, some special quasi-solvable models which
can be constructed from the $\Lsl(2)$ generators \cite{Turbi1} were
found to be type A $\cN$-fold supersymmetric \cite{AKOSW2,ASTY,AST1,
KlPl2,KlPl3,DoDuTa1,DoDuTa2}. Then, it was shown in Ref.~\cite{ANST1}
that the type A $\cN$-fold supersymmetric model is essentially equivalent
to the $\Lsl(2)$ quasi-solvable model. In this section, we will
review the equivalence and then clarify the relation between the pair
of the Hamiltonian $H_{\cN}^{\pm}$ in the framework of the $\Lsl(2)$
quasi-solvable models, which has not been discussed yet in the previous
articles. Let us first construct $H_{\cN}^{-}$ so that it is
quasi-solvable with respect to the type A operator (\ref{eqn:typea}).
The quasi-solvability condition (\ref{eqn:qscon}) in this case is,
\begin{align}
P_{\cN}H_{\cN}^{-}\cV_{\cN}^{-}=0,
\label{eqn:qscd1}
\end{align}
where $\cV_{\cN}^{-}$ is given by Eq.~(\ref{eqn:dfVn-}). On the
gauge-transformed space, the condition (\ref{eqn:qscd1}) is equivalent to,
\begin{align}
\tP_{\cN}\tH_{\cN}^{-}\tcV_{\cN}^{-}=0,
\label{eqn:qscd2}
\end{align}
where $\tP_{\cN}$ and $\tH_{\cN}^{-}$ are given by Eqs.~(\ref{eqns:gtPaH})
and $\tcV_{\cN}^{-}$ is defined by,
\begin{align}
\tcV_{\cN}^{-}=\ker\tP_{\cN}=\text{span}\,\left\{G_{\cN}U\phi :
 \phi\in\ker P_{\cN}\right\}.
\end{align}
Introducing a function $h(q)$ defined as a solution of the following
differential equation:
\begin{align}
h''(q)-E(q)h'(q)=0,
\label{eqn:defEq}
\end{align}
we find that $\tP_{\cN}$ is expressed in terms of $h$ as,
\begin{align}
\tP_{\cN}=\left(h'\right)^{\cN}\frac{d^{\cN}}{dh^{\cN}}.
\label{eqn:gtsch}
\end{align}
Thus, we easily have,
\begin{align}
\tcV_{\cN}^{-}=\text{span}\,\left\{1,h(q),\dots,h(q)^{\cN-1}\right\}.
\label{eqn:tsvsp}
\end{align}
From Eqs.~(\ref{eqn:qscd2}) and (\ref{eqn:tsvsp}), the quasi-solvability
condition for $\tH_{\cN}^{-}$ reads,
\begin{align}
\frac{d^{\cN}}{dh^{\cN}}\tH_{\cN}^{-}h^{k-1}=0\qquad\text{for}
 \qquad\forall\, k=1,\dots,\cN.
\label{eqn:qscd3}
\end{align}
Any constant is a trivial solution of (\ref{eqn:qscd3}). For $\cN=1$,
any first-order differential operator of the following form,
\begin{align}
f_{1}(h)\frac{d}{dh},
\label{eqn:deff1}
\end{align}
is a solution, while for $\cN\ge 2$ there are three independent
first-order differential operators as solutions of (\ref{eqn:qscd3}):
\begin{align}
\frac{d}{dh}&= J^{-}, & h\frac{d}{dh}&= J^{0}+\frac{\cN-1}{2},
 & h^{2}\frac{d}{dh}-(\cN-1)h&= J^{+}.
\label{eqn:sl2ge}
\end{align}
The $J^{+,0,-}$ defined above satisfy the $\Lsl(2)$ algebra:
\begin{align}
[J^{+},J^{-}]=-2J^{0},\qquad [J^{\pm},J^{0}]=\mp J^{\pm}.
\end{align}
In the same way, we find that for $\cN=1,2$, any second-order
differential operator of the following form,
\begin{align}
f_{2}(h)\frac{d^{2}}{dh^{2}},
\label{eqn:deff2}
\end{align}
is a solution, while for $\cN\ge 3$ there are five independent
second-order differential operators as solutions of (\ref{eqn:qscd3}):
\begin{gather*}
\frac{d^{2}}{dh^{2}}=\left(J^{-}\right)^{2},\qquad
h\frac{d^{2}}{dh^{2}}=J^{0}J^{-}+\frac{\cN-1}{2}J^{-},\\
h^{2}\frac{d^{2}}{dh^{2}}
 =\left(J^{0}\right)^{2}+(\cN-2)J^{0}+\frac{(\cN-1)(\cN-3)}{4},\\
h^{3}\frac{d^{2}}{dh^{2}}-(\cN-1)(\cN-2)h
 =J^{+}J^{0}+\frac{3\cN-5}{2}J^{+},\\
h^{4}\frac{d^{2}}{dh^{2}}-2(\cN-2)h^{3}\frac{d}{dh}
 +(\cN-1)(\cN-2)h^{2}=\left(J^{+}\right)^{2}.
\end{gather*}
Therefore, the general solution of (\ref{eqn:qscd3}) for $\cN\ge 3$
can be expressed as,
\begin{align}
\tH_{\cN}^{-}=-\sum_{\substack{i,j=+,0,-\\ i\ge j}}a_{ij}^{(-)}J^{i}J^{j}
 +\sum_{i=+,0,-}b_{i}^{(-)}J^{i}-C^{(-)},
\label{eqn:gsolH}
\end{align}
where $a_{ij}^{(-)}$, $b_{i}^{(-)}$ and $C^{(-)}$ are constants. This
gauged Hamiltonian is nothing but the $\Lsl(2)$ quasi-solvable model
\cite{Turbi1}.
We can set $a_{+-}^{(-)}=0$ without any loss of generality due to the
relation:
\begin{align}
\frac{1}{2}(J^{+}J^{-}+J^{-}J^{+})-(J^{0})^{2}=-\frac{1}{4}(\cN^{2}-1).
\end{align}
For convenience, we introduce new parameters as follows:
\begin{subequations}
\begin{gather}
a_{4}=a_{++}^{(-)},\quad a_{3}=a_{+0}^{(-)},\quad a_{2}=a_{00}^{(-)},
 \quad a_{1}=a_{0-}^{(-)},\quad a_{0}=a_{--}^{(-)},\\
b_{2}=-b_{+}^{(-)}-\frac{a_{+0}^{(-)}}{2},\quad b_{1}=-b_{0}^{(-)},
 \quad b_{0}=-b_{-}^{(-)}-\frac{a_{0-}^{(-)}}{2}.
\end{gather}
\end{subequations}
Then, the gauged Hamiltonian (\ref{eqn:gsolH}) is expressed in terms of
$h$ as,
\begin{align}
\tH_{\cN}^{-}&=-P(h)\frac{d^{2}}{dh^{2}}-\left[Q(h)-\frac{\cN-2}{2}
 P'(h)\right]\frac{d}{dh}\notag\\
&\equal -\left[R-\frac{\cN-1}{2}Q'(h)
 +\frac{(\cN-1)(\cN-2)}{12}P''(h)\right],
\label{eqn:gaugH}
\end{align}
where $P(h)$ and $Q(h)$ are given by,
\begin{align}
P(h)&=a_{4}h^{4}+a_{3}h^{3}+a_{2}h^{2}+a_{1}h+a_{0},
\label{eqn:defPh}
\\
Q(h)&=b_{2}h^{2}+b_{1}h+b_{0},
\label{eqn:defQh}
\end{align}
while $R=C^{(-)}+(\cN^{2}-1)a_{2}/12$ is a constant.
If the Hamiltonian (\ref{eqn:gaugH}) is gauge-transformed back to
the original Hamiltonian $H_{\cN}^{-}$ with the gauge factor $G_{\cN}U$,
it is in general not of the canonical form of the Schr\"{o}dinger
operator like Eq.~(\ref{eqn:ordiH}).
We can find that the original Hamiltonian $H_{\cN}^{-}$ becomes of the
canonical form if and only if the following conditions hold:
\begin{subequations}
\label{eqns:cano-}
\begin{align}
P(h)&=\frac{1}{2}\bigl(h'(q)\bigr)^{2},
\label{eqn:cano1}\\
Q(h)&=-W(q)h'(q).
\label{eqn:cano2}
\end{align}
\end{subequations}
Under the above conditions satisfied, we have,
\begin{align}
H_{\cN}^{-}=(G_{\cN}U)^{-1}\tH_{\cN}^{-}(G_{\cN}U)
 =-\frac{1}{2}\frac{d^{2}}{dq^{2}}+V_{\cN}^{-}(q),
\end{align}
where the potential $V_{\cN}^{-}$ is given by,
\begin{align}
V_{\cN}^{-}=-\frac{1}{12P}\left[(\cN^{2}-1)\left(PP''-\frac{3}{4}
 (P')^{2}\right)+3\cN (P'Q-2PQ')-3Q^{2}\right]-R.
\label{eqn:V-iPQ}
\end{align}
Substituting Eqs.~(\ref{eqn:defEq}) and (\ref{eqns:cano-}) for
Eq.~(\ref{eqn:V-iPQ}), we can check that the above
expression (\ref{eqn:V-iPQ}) for $V_{\cN}^{-}$ is identical with the one
in Eq.~(\ref{eqn:acond1}).

When $\cN=2$, any second-order operator (\ref{eqn:deff2}) can be added
to the gauged Hamiltonian (\ref{eqn:gsolH}) and thus all the $a_{ij}$
in Eq.~(\ref{eqn:gsolH}) can be arbitrary functions of $h$. As a
consequence, $P(h)$ in Eq.~(\ref{eqn:gaugH}) can be an arbitrary
function of $h$.

When $\cN=1$, any first-order operator (\ref{eqn:deff1}) in addition
to any second-order one (\ref{eqn:deff2}) can be added to the gauged
Hamiltonian (\ref{eqn:gsolH}) and thus all the $b_{i}$ in addition to
all the $a_{ij}$ in Eq.~(\ref{eqn:gsolH}) can be arbitrary functions
of $h$. As a consequence, both $P(h)$ and $Q(h)$ in Eq.~(\ref{eqn:gaugH})
can be arbitrary functions of $h$.

Combining the above considerations for $\cN=1,2$ with the results
(\ref{eqn:defPh}) and (\ref{eqn:defQh}) for $\cN\ge 3$, we obtain
the following conditions for $P(h)$ and $Q(h)$:
\begin{align}
\frac{d^{3}}{dh^{3}}Q(h)&=0\qquad\text{for}\qquad\cN\ge 2,
\label{eqn:condQ}
\\
\frac{d^{5}}{dh^{5}}P(h)&=0\qquad\text{for}\qquad\cN\ge 3.
\label{eqn:condP}
\end{align}
If we rewrite the above conditions in terms of $q$ with the aid of
Eqs.~(\ref{eqn:defEq}) and (\ref{eqns:cano-}),
we find that the condition (\ref{eqn:condQ}) is equivalent to
Eq.~(\ref{eqn:acond2}) while the condition (\ref{eqn:condP}) is
equivalent to Eq.~(\ref{eqn:acond3}). Therefore, the complete
correspondence between type A $\cN$-fold supersymmetric models and
the $\Lsl(2)$ quasi-solvable models has been established except for
the other Hamiltonian $H_{\cN}^{+}$, which will be discussed in the
next.

The partner Hamiltonian $H_{\cN}^{+}$ should be constructed such that
it is quasi-solvable with respect to $P_{\cN}^{\dagger}$. It is easy
to see that the operator $P_{\cN}^{\dagger}$ can be converted into
the form identical with $\tP_{\cN}$ by a gauge transformation with
another gauge factor $G_{\cN}U^{-1}$:
\begin{align}
\bar{P}_{\cN}^{\dagger}=i^{\cN}\left(G_{\cN}U^{-1}\right)
 P_{\cN}^{\dagger}\left(G_{\cN}U^{-1}\right)^{-1}
 =\left(h'\right)^{\cN}\frac{d^{\cN}}{dh^{\cN}}.
\end{align}
where $h$ is the same function of $q$ as defined previously by
Eq.~(\ref{eqn:defEq}).
In the above and hereafter, we attach bars to operators, vectors and
vector spaces to indicate that they are quantities gauge-transformed
with the gauge factor $G_{\cN}U^{-1}$. Therefore, the quasi-solvability
condition for the gauged Hamiltonian $\bar{H}_{\cN}^{+}$ is completely
the same as that for $\tH_{\cN}^{-}$, Eq.~(\ref{eqn:qscd3}).
This means that $\bar{H}_{\cN}^{+}$ is another quasi-solvable model
constructed from the $\Lsl(2)$ generators (\ref{eqn:sl2ge}) and thus
has the same form as Eqs.~(\ref{eqn:gsolH}) and (\ref{eqn:gaugH}):
\begin{subequations}
\begin{align}
\bar{H}_{\cN}^{+}
&=-\sum_{\substack{i,j=+,0,- \\ i\ge j}}a_{ij}^{(+)}J^{i}J^{j}
 +\sum_{i=+,0,-}b_{i}^{(+)}J^{i}-C^{(+)}
\label{eqn:algH+}\\
&=-P^{+}(h)\frac{d^{2}}{dh^{2}}-\left[Q^{+}(h)-\frac{\cN-2}{2}
 P^{+'}(h)\right]\frac{d}{dh}\notag\\
&\equal -\left[R^{+}-\frac{\cN-1}{2}Q^{+'}(h)
 +\frac{(\cN-1)(\cN-2)}{12}P^{+''}(h)\right],
\label{eqn:gauH+}
\end{align}
\end{subequations}
where $a_{ij}^{(+)}$, $b_{i}^{(+)}$, $C^{(+)}$ and $R^{+}$ are constants,
and $P^{+}(h)$ and $Q^{+}(h)$ is a polynomial of fourth- and
second-degree, respectively. Up to now, there is no relation between
$\tH_{\cN}^{-}$ and $\bar{H}_{\cN}^{+}$. If we demand that
$\bar{H}_{\cN}^{+}$ be gauge-transformed, with the gauge factor
$G_{\cN}U^{-1}$, back to an operator of the Schr\"{o}dinger form,
we have,
\begin{align}
P^{+}(h)=\frac{1}{2}\bigl(h'(q)\bigr)^{2},\qquad Q^{+}(h)=W(q)h'(q).
\end{align}
Combining the above with Eq.~(\ref{eqns:cano-}), we yield the following
relations:
\begin{align}
P^{+}(h)=P(h),\qquad Q^{+}(h)=-Q(h).
\label{eqn:relPQ}
\end{align}
Then, the partner Hamiltonian $H_{\cN}^{+}$ becomes,
\begin{align}
H_{\cN}^{+}=\left(G_{\cN}U^{-1}\right)^{-1}\bar{H}_{\cN}^{+}\left(
 G_{\cN}U^{-1}\right)=-\frac{1}{2}\frac{d^{2}}{dq^{2}}+V_{\cN}^{+}(q),
\end{align}
with,
\begin{align}
V_{\cN}^{+}=-\frac{1}{12P}\left[(\cN^{2}-1)\left( PP''-\frac{3}{4}
 (P')^{2}\right)-3\cN (P'Q-2PQ')-3Q^{2}\right]-R^{+}.
\label{eqn:V+iPQ}
\end{align}
To establish the relation between $R$ and $R^{+}$, we should recall
the formula \cite{AST2}: $H_{\cN}^{+}-H_{\cN}^{-}=iw'_{\cN-1}$, where
$w_{\cN-1}$ is defined in Eq.~(\ref{eqn:nfsch}). For the type A
$\cN$-fold supercharge (\ref{eqn:typea}), we have $w_{\cN-1}=-i\cN W$.
From Eqs.~(\ref{eqn:cano2}), (\ref{eqn:V-iPQ}) and (\ref{eqn:V+iPQ})
we finally obtain,
\begin{align}
R^{+}=R.
\label{eqn:relRR}
\end{align}
We can easily see that the potential (\ref{eqn:V+iPQ}) with
Eq.~(\ref{eqn:relRR}) is identical with $V_{\cN}^{+}$ in
Eq.~(\ref{eqn:acond1}). Thus, the models constructed in this section
are completely identical with the ones by the analytic construction
in the preceding section.

Before closing the section, we will refer to an interesting relation
between the algebraic Hamiltonians $\tH_{\cN}^{-}$ and
$\bar{H}_{\cN}^{+}$. From Eqs.~(\ref{eqn:relPQ}) and (\ref{eqn:relRR}),
the relation between the parameters in $\tH_{\cN}^{-}$ and the ones in
$\bar{H}_{\cN}^{+}$ reads,
\begin{subequations}
\label{eqns:relpa}
\begin{gather}
a_{ij}^{(+)}=a_{ij}^{(-)}\quad (\forall\, i,j=+,0,-),\qquad
 C^{(+)}=C^{(-)},\\
b_{+}^{(+)}+b_{+}^{(-)}+a_{+0}^{(-)}=0,\qquad
 b_{0}^{(+)}+b_{0}^{(-)}=0,\qquad
 b_{-}^{(+)}+b_{-}^{(-)}+a_{0-}^{(-)}=0.
\end{gather}
\end{subequations}
Substituting the above for Eq.~(\ref{eqn:algH+}), we can rewrite
$\bar{H}_{\cN}^{+}$ in terms of $a_{ij}^{(-)}$ and $b_{i}^{(-)}$:
\begin{align}
\bar{H}_{\cN}^{+}=-\sum_{\substack{i,j=+,0,-\\i\ge j}}a_{ij}^{(-)}
 J^{j}J^{i}-\sum_{i=+,0,-}b_{i}^{(-)}J^{i}-C^{(-)}.
\label{eqn:psolH}
\end{align}
Comparing this expression with Eq.~(\ref{eqn:gsolH}), we see that
$\bar{H}_{\cN}^{+}$ is obtained from $\tH_{\cN}^{-}$ by interchanging
the order of the quadratic terms and by interchanging the sign of the
coefficients of the linear terms.

\section{\label{sec:class}Classification of Type A Models}

It was shown that the $\Lsl(2)$ quasi-solvable models can be classified
using the shape invariance of the Hamiltonian under the action of
$GL(2,K)$ ($K=\bbR$ or $\bbC$) of linear fractional transformations
\cite{LoKaOl3,LoKaOl4}. The equivalence established in the previous
section ensures that the type A $\cN$-fold supersymmetric Hamiltonians
can be fit into the same classification scheme. Then, a natural question
arises whether the type A $\cN$-fold supercharges can be also classified
in the same scheme or not. This question was left as an open problem
in the previous paper \cite{ANST2} and will be completely answered in
this section.

\subsection{\label{ssec:gl2ksi}$GL(2,K)$ Shape Invariance}

The linear fractional transformation of $h$ is introduced by,
\begin{align}
h\mapsto \hat{h}=\frac{\alpha h+\beta}{\gamma h+\delta}\qquad
 (\alpha,\beta,\gamma,\delta\in K;\ \Delta =\alpha\delta
 -\beta\gamma\neq 0).
\label{eqn:lftoh}
\end{align}
Then, it turns out that the gauged Hamiltonian (\ref{eqn:gaugH}) is
shape invariant under the following transformation induced by
Eq.~(\ref{eqn:lftoh}):
\begin{align}
\tH_{\cN}^{-}(h)\mapsto\widehat{\tH}{}_{\cN}^{-}(h)=(\gamma h
 +\delta)^{\cN-1}\tH_{\cN}^{-}(\hat{h})(\gamma h+\delta)^{-(\cN-1)},
\end{align}
where $P(h)$ and $Q(h)$ in Eq.~(\ref{eqn:gaugH}) are transformed
according to,
\begin{subequations}
\label{eqns:tfPaQ}
\begin{align}
P(h)&\mapsto\hat{P}(h)=\Delta^{-2}(\gamma h+\delta)^{4}P(\hat{h}),
\label{eqn:tfofP}\\
Q(h)&\mapsto\hat{Q}(h)=\Delta^{-1}(\gamma h+\delta)^{2}Q(\hat{h}).
\end{align}
\end{subequations}

\subsection{\label{ssec:invham}Invariance of the Hamiltonian}

In the next, we will examine the transformation of the original
Hamiltonian $H_{\cN}^{\pm}$. Since the function $h(q)$ is determined by
Eq.~(\ref{eqn:cano1}), we have,
\begin{align}
|q|=\Int\frac{dh}{\sqrt{2P(h)}}\mapsto |q|=\Int
 \frac{dh}{\sqrt{2\hat{P}(h)}}
&=\Int dh\frac{\Delta}{(\gamma h+\delta)^{2}}\frac{1}{\sqrt{2P(\hat{h})}}
\notag\\
&=\Int\frac{d\hat{h}}{\sqrt{2P(\hat{h})}}.
\label{eqn:ellip}
\end{align}
Therefore, $h(q)$ before the transformation is identical with $\hat{h}(q)$
as a function of $q$. In other words, the relation between $h$ before
the transformation (denoted by $h_{old}$) and $h$ after the transformation
(denoted by $h_{new}$) is, as a function of $q$, consistent with
Eq.~(\ref{eqn:lftoh}):
\begin{align}
h_{old}(q)=\hat{h}(q)
 =\frac{\alpha\, h_{new}(q)+\beta}{\gamma\, h_{new}(q)+\delta}.
\label{eqn:tfhoq}
\end{align}
On the other hand, the potentials $V_{\cN}^{\pm}$ can be rewritten as
\cite{LoKaOl3},
\begin{align}
V_{\cN}^{\pm}=-\frac{1}{24P}\left[2(\cN^{2}-1)H[P]\mp 3\cN (P,Q)^{(1)}
 -6Q^{2}\right]-R,
\label{eqn:invfV}
\end{align}
where $H[P]$ is the Hessian of $P$ and $(P,Q)^{(1)}$ are the first
transvectant of $P$ and $Q$, given by (see also Appendix~\ref{app:invsp}),
\begin{align}
H(P)=P P''-\frac{3}{4}\left(P'\right)^{2},\qquad
 (P,Q)^{(1)}=2P' Q-4PQ'.
\label{eqn:Hessi}
\end{align}
All the objects $O(h)$ in both the numerator and denominator of
Eq.~(\ref{eqn:invfV}), namely, $H[P]$, $(P,Q)^{(1)}$, $Q^{2}$ and
$P$ transform according to,
\begin{align}
O(h_{old})\mapsto \hat{O}(h_{new})
 =\Delta^{-2}(\gamma h_{new}+\delta)^{4}O(\hat{h}),
\end{align}
that is, they belong to the multiplier representation $\rho_{4,\, -2}$
of $GL(2,K)$ defined in Eq.~(\ref{eqn:mlrep}).
As a consequence, the functional form of $V_{\cN}^{\pm}$ are preserved
under the transformation:
\begin{align}
V_{\cN}^{\pm}(h_{old})\mapsto
 \hat{V}_{\cN}^{\pm}(h_{new})=V_{\cN}^{\pm}(\hat{h}).
\end{align}
Finally, since the functional form of $h_{old}(q)$ and $\hat{h}(q)$ is
identical with each other, Eq.~(\ref{eqn:tfhoq}), the potential
$V_{\cN}^{\pm}$ is invariant as functions of $q$:
\begin{align}
V_{\cN}^{\pm}\bigl(h_{old}(q)\bigr)\mapsto\hat{V}_{\cN}^{\pm}
 \bigl(h_{new}(q)\bigr)=V_{\cN}^{\pm}\bigl(\hat{h}(q)\bigr)
 =V_{\cN}^{\pm}\bigr(h_{old}(q)\bigr).
\end{align}
Therefore, the type A $\cN$-fold supersymmetric Hamiltonians
$H_{\cN}^{\pm}$ are invariant under the $GL(2,K)$ transformation.

\subsection{\label{ssec:invsch}Invariance of the Supercharge}

In the next, we will examine the transformation of the type A
$\cN$-fold supercharge $P_{\cN}$. The gauged $\cN$-fold
supercharge (\ref{eqn:gtsch}) is transformed according to,
\begin{align}
\tP_{\cN}=(h'_{old})^{\cN}\frac{d^{\cN}}{dh_{old}^{\cN}}
 \mapsto\widehat{\tP}_{\cN}=(\gamma h_{new}+\delta)^{\cN-1}
 (\hat{h}')^{\cN}\frac{d^{\cN}}{d\hat{h}^{\cN}}
 (\gamma h_{new}+\delta)^{-(\cN-1)}.
\label{eqn:tPtfs}
\end{align}
On the other hand, the gauge factors $G_{\cN}U^{\pm 1}$ are transformed as,
\begin{align}
G_{\cN}U^{\pm 1}(h_{old})&=\exp\left[\Int dq\left(\frac{\cN-1}{2}E\pm W
 \right)\right]=\exp\left[\Int dh_{old}\frac{(\cN-1)P'(h_{old})\mp
 2Q(h_{old})}{4P(h_{old})}\right]\notag\\
&\mapsto\widehat{G_{\cN}U^{\pm 1}}(h_{new})=\exp\left[\Int dh_{new}
 \frac{(\cN-1)\hat{P}'(h_{new})\mp 2\hat{Q}(h_{new})}{4\hat{P}(h_{new})}
 \right].
\label{eqn:GUtf1}
\end{align}
From Eqs.~(\ref{eqn:lftoh}) and (\ref{eqns:tfPaQ}), the r.h.s. of
Eq.~(\ref{eqn:GUtf1}) is calculated as,
\begin{align}
\widehat{G_{\cN}U^{\pm 1}}(h_{new})&=\exp\left[\Int d\hat{h}\frac{(\cN-1)
 P'(\hat{h})\mp 2Q(\hat{h})}{4P(\hat{h})}+\Int dh_{new}\frac{(\cN-1)
 \gamma}{\gamma h_{new}+\delta}\right]\notag\\
&=(\gamma h_{new}+\delta)^{\cN-1}G_{\cN}U^{\pm 1}(\hat{h}).
\label{eqn:GUtf2}
\end{align}
The $\cN$-fold supercharge $P_{\cN}$ is expressed as,
\begin{align}
P_{\cN}&=(-i)^{\cN}\bigl(G_{\cN}U(h_{old})\bigr)^{-1}\tP_{\cN}
 \bigl(G_{\cN}U(h_{old})\bigr)\notag\\
&=(-ih'_{old})^{\cN}\bigl(G_{\cN}U(h_{old})\bigr)^{-1}
 \frac{d^{\cN}}{dh_{old}^{\cN}}\bigl(G_{\cN}U(h_{old})\bigr).
\label{eqn:schoh}
\end{align}
Combining Eqs.~(\ref{eqn:tPtfs})--(\ref{eqn:GUtf2}), we obtain,
\begin{align}
P_{\cN}\mapsto \hat{P}_{\cN}&=
 (-i)^{\cN}\bigl(\widehat{G_{\cN}U}(h_{new})\bigr)^{-1}\widehat{\tP}_{\cN}
 \bigl(\widehat{G_{\cN}U}(h_{new})\bigr)\notag\\
&=(-i\hat{h}')^{\cN}\bigl(G_{\cN}U(\hat{h})\bigr)^{-1}
 \frac{d^{\cN}}{d\hat{h}^{\cN}}\bigl(G_{\cN}U(\hat{h})\bigr).
\label{eqn:tfsch}
\end{align}
Comparing Eq.~(\ref{eqn:schoh}) with (\ref{eqn:tfsch}), we see that
$\hat{P}_{\cN}$ is obtained from $P_{\cN}$ with $h_{old}(q)$ replaced
by $\hat{h}(q)$. From the fact that $\hat{h}(q)$ is identical with
$h_{old}(q)$ as a function of $q$, Eq.~(\ref{eqn:tfhoq}), we
finally conclude,
\begin{align}
P_{\cN}=\hat{P}_{\cN},
\end{align}
that is, the $\cN$-fold supercharge is also invariant under the $GL(2,K)$
transformation. We note that the manifest invariance of the $\cN$-fold
supercharge is lost if we express it in terms of $E(q)$ and $W(q)$ as
Eq.~(\ref{eqn:typea}). This is because $E(q)$ is not an invariant
function under the $GL(2,K)$ transformation, as we will see below.
From Eq.~(\ref{eqn:tfhoq}), we have,
\begin{align}
\frac{h''_{old}(q)}{h'_{old}(q)}=
 \frac{\hat{h}''(q)}{\hat{h}'(q)}=\frac{h''_{new}(q)}{h'_{new}(q)}
 -\frac{2\gamma\, h'_{new}(q)}{\gamma\, h_{new}(q)+\delta}.
\end{align}
The function $h(q)$ is defined so that Eq.~(\ref{eqn:defEq}) is
fulfilled and thus the relation between $E(q)$ before and after the
transformation reads,
\begin{align}
E(q)=\hat{E}(q)-\frac{2\gamma\, h'_{new}(q)}{\gamma\, h_{new}(q)+\delta}.
\end{align}
On the other hand, $W(q)$ is an invariant function:
\begin{align}
\hat{W}(q)=-\frac{\hat{Q}\bigl(h_{new}(q)\bigr)}{h'_{new}(q)}
 =-\frac{Q\bigl(\hat{h}(q)\bigr)}{\hat{h}'(q)}=W(q).
\end{align}
Therefore, the $\cN$-fold supercharge of the form (\ref{eqn:typea})
is expressed as,
\begin{align}
P_{\cN}&=(-i)^{\cN}\prod_{k=-(\cN-1)/2}^{(\cN-1)/2}\left(\frac{d}{dq}+W(q)
 -kE(q)\right)\notag\\
&=(-i)^{\cN}\prod_{k=-(\cN-1)/2}^{(\cN-1)/2}\left(\frac{d}{dq}+\hat{W}(q)
 -k \hat{E}(q)+k \frac{2\gamma\, h'_{new}(q)}{\gamma\, h_{new}(q)
 +\delta}\right).
\end{align}

\subsection{\label{ssec:exampl}An example}

As an example, we will demonstrate the equivalence between the case
$P(h)=2h$ and $P(h)=2h^{3}$. In the previous paper \cite{ANST2}, we
classified them as different cases, namely, case (1) for the former
and case (3a) for the latter. However, it is easy to see that the latter
is obtained from the former by the following $GL(2,K)$ transformation:
\begin{align}
h_{old}\mapsto\hat{h}=\frac{1}{h_{new}}.
\label{eqn:tfhex}
\end{align}
In this case, the transformation of $P$ and $Q$ defined by
Eqs.~(\ref{eqns:tfPaQ}) reads,
\begin{alignat}{3}
&\equal  &\qquad P(h_{old})&=2h_{old}, &\qquad
 Q(h_{old})&=b_{2}h_{old}^{2}+b_{1}h_{old}+b_{0}\notag\\
&\mapsto & \hat{P}(h_{new})&=2h_{new}^{3}, &
 \hat{Q}(h_{new})&=-b_{0}h_{new}^{2}-b_{1}h_{new}-b_{2}.
\label{eqn:PaQex}
\end{alignat}
From Eq.~(\ref{eqn:ellip}), the functions $h_{old}(q)$ and $h_{new}(q)$
are calculated as,
\begin{align}
h_{old}(q)=q^{2},\qquad h_{new}(q)=\frac{1}{q^{2}},
\label{eqn:hoqex}
\end{align}
which is consistent with Eq.~(\ref{eqn:tfhex}). The transvectants
appeared in Eq.~(\ref{eqn:invfV}) become,
\begin{alignat}{3}
&\equal  &\qquad H(P)&=-3, &\qquad
 (P,Q)^{(1)}&=-4\left(3b_{2}h_{old}^{2}+b_{1}h_{old}-b_{0}\right)\notag\\
&\mapsto & H(\hat{P})&=-3h_{new}^{4}, &
 (\hat{P},\hat{Q})^{(1)}&=4\left(b_{0}h_{new}^{4}-b_{1}h_{new}^{3}
 -3b_{2}h_{new}^{2}\right).
\label{eqn:Hesex}
\end{alignat}
Substituting Eqs.~(\ref{eqn:PaQex}) and (\ref{eqn:Hesex}) for
Eq.~(\ref{eqn:invfV}), we have the potentials $V_{\cN}^{\pm}$ and
$\hat{V}_{\cN}^{\pm}$ as functions of $h$:
\begin{subequations}
\label{eqns:potex}
\begin{align}
V_{\cN}^{\pm}(h_{old})&=\frac{1}{8}h_{old}(b_{2}h_{old}+b_{1})^{2}
 +\frac{b_{2}}{4}(b_{0}\mp 3\cN)h_{old}\notag\\
&\equal +\frac{b_{0}^{2}\pm 2\cN b_{0}+\cN^{2}-1}{8h_{old}}
 +\frac{b_{1}}{4}(b_{0}\mp\cN)-R,\\
\hat{V}_{\cN}^{\pm}(h_{new})&=\frac{1}{8h_{new}}\left(
 \frac{b_{2}}{h_{new}}+b_{1}\right)^{2}+\frac{b_{2}(b_{0}\mp 3\cN)}{
 4h_{new}}\notag\\
&\equal +\frac{b_{0}^{2}\pm 2\cN b_{0}+\cN^{2}-1}{8}h_{new}
 +\frac{b_{1}}{4}(b_{0}\mp\cN)-R.
\end{align}
\end{subequations}
Finally, we confirm the invariance of the potentials as functions of $q$
by substituting Eq.~(\ref{eqn:hoqex}) for the above (\ref{eqns:potex}):
\begin{align}
V_{\cN}^{\pm}\bigl(h_{old}(q)\bigr)=\hat{V}_{\cN}^{\pm}\bigl(h_{new}(q)
 \bigr)&=\frac{1}{8}q^{2}\left(b_{2}q^{2}+b_{1}\right)^{2}+\frac{b_{2}}{4}
 (b_{0}\mp 3\cN)q^{2}\notag\\
&\equal +\frac{b_{0}^{2}\pm 2\cN b_{0}+\cN^{2}-1}{8q^{2}}
 +\frac{b_{1}}{4}(b_{0}\mp\cN)-R.
\end{align}
The functions $E(q)$ and $W(q)$ are calculated as,
\begin{align}
E(q)=\frac{h''_{old}(q)}{h'_{old}(q)}=\frac{1}{q},\qquad
 \hat{E}(q)=\frac{h''_{new}(q)}{h'_{new}(q)}=-\frac{3}{q},\\
W(q)=\hat{W}(q)=-\frac{1}{2}\left(b_{2}q^{3}+b_{1}q+\frac{b_{0}}{q}\right).
\end{align}
Thus, the type A $\cN$-fold supercharge of the form (\ref{eqn:typea})
reads,
\begin{align}
P_{\cN}=\hat{P}_{\cN}=(-i)^{\cN}\prod_{k=-(\cN-1)/2}^{(\cN-1)/2}\left(
 \frac{d}{dq}-\frac{b_{2}}{2}q^{3}-\frac{b_{1}}{2}q-\frac{b_{0}}{2q}
 +\frac{k}{q}\right).
\end{align}

\subsection{\label{ssec:classi}Classification of Models}

For a given $P(h)$, the function $h(q)$ is determined by
Eq.~(\ref{eqn:ellip}) and a particular type A $\cN$-fold supersymmetric
model is obtained by substituting this $h(q)$ for Eq.~(\ref{eqn:invfV}).
Since the potentials $V_{\cN}^{\pm}(q)$ and the $\cN$-fold supercharge
are invariant under the $GL(2,K)$ transformation, the type A models
can be classified according to the inequivalent elliptic integral
(\ref{eqn:ellip}) under the transformation.
The elliptic integral (\ref{eqn:ellip}) can be classified according
to the distribution of the zeros of $P(h)$, e.g., multiplicity of
the zeros.
This idea was first introduced in Ref.~\cite{LoKaOl3} to classify
the $\Lsl(2)$ quasi-solvable models.
Under the transformation (\ref{eqn:tfofP}) of $GL(2,\bbR)$ or
$GL(2,\bbC)$, every quartic polynomial $P(h)$ with real or complex
coefficients is equivalent to one of the eight or five forms,
respectively, as shown in Table~\ref{tab:class}.
\begin{table}[h]
\begin{center}
\[
\arraycolsep=5mm
\begin{array}{|c|c|c|}\hline
\text{Case} & GL(2,\bbR) & GL(2,\bbC) \\
\hline
\text{I} & 1/2 &1/2 \\
\hline
\text{II} & 2h & 2h \\
\hline
\text{III} & 2\nu h^{2} & 2\nu h^{2} \\
\text{III$'$} & \nu (h^{2}+1)^{2}/2 & \\
\hline
\text{IV} & 2\nu (h^{2}-1) & 2\nu (h^{2}-1) \\
\text{IV$'$} & 2\nu(h^{2}+1) & \\
\hline
\text{V} & 2h^{3}-g_{2}h/2-g_{3}/2  & 2h^{3}-g_{2}h/2-g_{3}/2 \\
\text{V$'$} & \nu (h^{2}+1)[(1-k^{2})h^{2}+1]/2 & \\
\hline
\end{array}
\]
\caption{The representatives of $P(h)$ under the $GL(2,\bbR)$ and
$GL(2,\bbC)$ transformations.}
\label{tab:class}
\end{center}
\end{table}
In Table~\ref{tab:class}, $\nu,g_{2},g_{3}\in K$ according to the
transformation group $GL(2,K)$, and $\nu\neq 0$, $0<k<1$,
$g_{2}^{3}-27g_{3}^{2}\neq 0$. In Ref.~\cite{Tanak1,Tanak2}, more general
quasi-solvable $M$-body systems constructed by $\Lsl(M+1)$ generators
are classified according to the above scheme and the explicit form of
the potential for each the cases is shown. The type A models in this
article correspond to the models for $M=1$ (with $b_{i}\rightarrow
\pm b_{i}$ for $H_{\cN}^{\mp}$) in Ref.~\cite{Tanak1,Tanak2}.
So, we do not repeat the exhibition of the potentials in this article.

\section{\label{sec:2f2dp}2-fold and 2nd-derivative Polynomial
Supersymmetry}

All that we have not investigated yet on type A $\cN$-fold supersymmetry
is the anti-commutator of $Q_{\cN}^{\dagger}$ and $Q_{\cN}$, namely,
the mother Hamiltonian. For this purpose, it is quite instructive to
analyze 2-fold supersymmetry. Let us first consider general 2-fold
supersymmetry, in which 2-fold supercharge is given by, 
\begin{align}
P_{2}=p^{2}+w_{1}(q)p+w_{0}(q).
\label{eqn:2fsc}
\end{align}
In this case, a pair of Hamiltonians $H_{2}^{\pm}=p^{2}/2+V_{2}^{\pm}(q)$
satisfies the intertwining relation $P_{2}H_{2}^{-}-H_{2}^{+}P_{2}=0$
if and only if,
\begin{subequations}
\label{eqns:2fcon}
\begin{align}
V_{2}^{\pm}(q)&=-\frac{1}{8}w_{1}(q)^{2}+\frac{1}{4}\left[
 \frac{w_{1}''(q)}{w_{1}(q)}-\frac{w_{1}'(q)^{2}}{2w_{1}(q)^{2}}
 +\frac{2C_{1}}{w_{1}(q)^{2}}\right]\pm\frac{i}{2}w_{1}'(q)-C_{2},
\label{eqn:2fco1}\\
w_{0}(q)&=\frac{1}{4}w_{1}(q)^{2}+\frac{1}{2}\left[
 \frac{w_{1}''(q)}{w_{1}(q)}-\frac{w_{1}'(q)^{2}}{2w_{1}(q)^{2}}
 +\frac{2C_{1}}{w_{1}(q)^{2}}\right]-\frac{i}{2}w_{1}'(q).
\label{eqn:2fco2}
\end{align}
\end{subequations}
where $C_{i}$ are arbitrary constants. From the form of $V_{2}^{\pm}(q)$,
it is indicated that $w_{1}^{*}(q)=-w_{1}(q)$, that is, $w_{1}(q)$ is
pure imaginary in order that $V_{2}^{\pm}(q)$ be real.
The above result was first reported in Refs.~\cite{AnIoCaDe1,AnIoNi1}
and later reexamined in Ref.~\cite{AST2}.
In Refs.~\cite{AnIoCaDe1,AnIoNi1}, the anti-commutator of the 2-fold
supercharges was given by the following form:
\begin{align}
2\cH_{2}=\{Q_{2}^{\dagger},Q_{2}\}
 =4\left(\bH_{2}+C_{2}\right)^{2}+C_{1},
\label{eqn:mhag2}
\end{align}
that is, it is a polynomial of degree 2 in $\bH_{2}$. Later, it was
proved in Ref.~\cite{AST1} that, for all $\cN$, the anti-commutator
of $Q_{\cN}^{\dagger}$ and $Q_{\cN}$ becomes a polynomial of degree $\cN$
in the original Hamiltonian $\bH_{\cN}$ if $Q_{\cN}^{\dagger}$ and
$Q_{\cN}$ satisfying Eq.~(\ref{eqn:nfalg2}) are uniquely determined
for the given $\bH_{\cN}$. Furthermore, it was also proved in
Ref.~\cite{AST1} that, for the above 2-fold supersymmetric systems,
2-fold supercharges are unique unless there is a constant $C_{3}$ which
satisfies,
\begin{align}
w_{1}''(q)-2i\, w_{1}(q)w_{1}'(q)-2i\, V_{2}^{-'}(q)=2C_{3}w_{1}'(q),
\end{align}
or equivalently,
\begin{align}
w_{1}'(q)-i\bigl(w_{1}(q)\bigr)^{2}-2i\, V_{2}^{-}(q)=2C_{3}w_{1}(q),
\label{eqn:nuni2}
\end{align}
where the integral constant is omitted since it can be absorbed in
$V_{2}^{-}(q)$. If Eq.~(\ref{eqn:nuni2}) is the case, the 2-fold
supersymmetric Hamiltonians $H_{2}^{\pm}$ satisfy another intertwining
relation $\iD P_{2}H_{2}^{-}-H_{2}^{+}\iD P_{2}=0$ and its conjugation
with the following 1-fold supercharges:
\begin{align}
\iD P_{2}=p+w_{1}(q)-iC_{3},\qquad
 \iD P_{2}^{\dagger}=p-w_{1}(q)+iC_{3}.
\end{align}
This result is almost evident from the fact that the condition
(\ref{eqn:nuni2}) together with Eq.~(\ref{eqn:2fco1}) implies,
\begin{align}
V_{2}^{\pm}(q)=\frac{1}{2}\bigl(iw_{1}(q)+C_{3}\bigr)^{2}\pm
 \frac{i}{2}w_{1}'(q)-\frac{C_{3}^{2}}{2},
\end{align}
that is, the 2-fold supersymmetric Hamiltonian $\bH_{2}$ in this case
is simultaneously ordinary supersymmetric (except for the irrelevant
constant term),
\begin{align}
\bigl[\iD Q_{2}, \bH_{2}\bigr]=\bigl[\iD Q_{2}^{\dagger},
 \bH_{2}\bigr]=0,\qquad
\bigl\{\iD Q_{2}^{\dagger}, \iD Q_{2}\bigr\}=2\bH_{2}+C_{3}^{2},
\end{align}
with respect to the supercharges defined by,
\begin{align}
\iD Q_{2}=\iD P_{2}^{\dagger}\,\psi,\qquad
 \iD Q_{2}^{\dagger}=\iD P_{2}\,\psi^{\dagger}.
\end{align}
\newpage\noindent
Therefore, if we define new 2-fold supercharges by,
\begin{subequations}
\label{eqns:df2sc}
\begin{align}
Q_{2}(\lambda)&=Q_{2}+\lambda^{*}\iD Q_{2}
 =\bigl(P_{2}^{\dagger}+\lambda^{*}\iD P_{2}^{\dagger}\bigr)\psi,\\
Q_{2}^{\dagger}(\lambda)&=Q_{2}^{\dagger}+\lambda\,\iD Q_{2}^{\dagger}
 =\bigl(P_{2}+\lambda\,\iD P_{2}\bigr)\psi^{\dagger},
\end{align}
\end{subequations}
the 2-fold supersymmetric Hamiltonian $\bH_{2}$ with $w_{1}(q)$
satisfying Eq.~(\ref{eqn:nuni2}) commutes with $Q_{2}(\lambda)$ and
$Q_{2}^{\dagger}(\lambda)$ for arbitrary $\lambda$:
\begin{align}
\bigl[Q_{2}(\lambda),\bH_{2}\bigr]
 =\bigl[Q_{2}^{\dagger}(\lambda),\bH_{2}\bigr]=0.
\end{align}
On the other hand, the anti-commutator of these new 2-fold supercharges
becomes,
\begin{align}
\bigl\{Q_{2}^{\dagger}(\lambda),Q_{2}(\lambda)\bigr\}
&=4\left(\bH_{2}+C_{2}\right)^{2}+|\lambda|^{2}\left(2\bH_{2}+C_{3}^{2}
 \right)+C_{1}\notag\\
&\equal +\lambda\,\bigl\{ Q_{2},\iD Q_{2}^{\dagger}\bigr\}
 +\lambda^{*}\bigl\{ Q_{2}^{\dagger},\iD Q_{2}\bigr\}.
\end{align}
It is now evident that the above anti-commutator cannot be, in general,
a polynomial in $\bH_{2}$ since it contains the term as
$(\lambda+\lambda^{*})p^{3}$. Therefore, 2-fold supersymmetry does not
always correspond to 2nd-derivative polynomial supersymmetry.

Next, let us return to the case of type A 2-fold supersymmetry.
From the definition of the type A $\cN$-fold supercharge (\ref{eqn:typea}),
type A 2-fold supersymmetry is a special case of 2-fold supersymmetry
where $w_{1}(q)$ and $w_{0}(q)$ in Eq.~(\ref{eqn:2fsc}) are given by,
\begin{align}
w_{1}(q)=-2iW(q),\qquad
 w_{0}(q)=-\left(W'(q)+W(q)^{2}+\frac{E'(q)}{2}-\frac{E(q)^{2}}{4}\right).
\label{eqn:re2A2}
\end{align}
Substituting the above for the condition (\ref{eqns:2fcon}), we see
that Eq.~(\ref{eqn:2fco1}) is equivalent to Eq.~(\ref{eqn:acond1})
while Eq.~(\ref{eqn:2fco2}) to Eq.~(\ref{eqn:acond2}). Furthermore,
if we substitute Eq.~(\ref{eqn:re2A2}) for Eq.~(\ref{eqn:nuni2}), we
find that the uniqueness of the type A 2-fold supercharge is guaranteed
unless there is a constant $C_{3}$ which satisfies,
\begin{align}
12W(q)^{2}-E(q)^{2}+2E'(q)=-16C_{3}W(q).
\label{eqn:nunA1}
\end{align}
In terms of $P(h)$ and $Q(h)$, the above condition is rewritten as,
\begin{align}
\left(3Q(h)^{2}+H[P]\right)^{2}=32C_{3}^{2}P(h)Q(h)^{2}.
\label{eqn:nunA2}
\end{align}
When $P(h)$ is (at most) a polynomial of fourth degree (remember that
$P(h)$ can be an arbitrary function in the 2-fold supersymmetric case),
both side of
Eq.~(\ref{eqn:nunA2}) are (at most) polynomials of eighth degree
belonging to the multiple representation $\rho_{\, 8,\,-4}$ defined
in Eq.~(\ref{eqn:mlrep}). Therefore, Eq.~(\ref{eqn:nunA2}) is a
polynomial identity and the set of its solution constitutes a hyperplane
$\Gamma$ in the parameter space $\bbR^{5+3}$ or $\bbC^{5+3}$ spanned by
$\{a_{i}, b_{i}\}$. On this hyperplane $\Gamma$, 2-fold supercharges are
not determined uniquely and the type A 2-fold supercharges can be deformed
as Eq.~(\ref{eqns:df2sc}) without destroying 2-fold supersymmetry.
However, it should be noted that the polynomiality of the anti-commutator
of the \textit{undeformed} type A 2-fold supercharges is preserved
on $\Gamma$.
The uniqueness of $\cN$-fold supercharges is only a sufficient but not
a necessary condition for the polynomiality. Conversely, we can choose
$\cN$-fold supercharges such that the anti-commutator of them becomes
a polynomial in the Hamiltonian.

We note that the general 2-fold supersymmetry is weakly quasi-solvable
since we cannot generally obtain two independent analytic elements of
$\ker P_{2}$ where $P_{2}$ is given by Eq.~(\ref{eqn:2fsc}) with
(\ref{eqn:2fco2}). Nevertheless, we can know the two spectra in the
solvable sector. The mother Hamiltonian (\ref{eqn:mhag2}) is always
a polynomial in the original Hamiltonian. Since it corresponds to
the characteristic polynomial which determines the spectra in the
solvable sector, they are given by the solutions of $4(E+C_{2})^{2}+C_{1}
=0$. This example shows a novel feature of weak quasi-solvability.
Even if, for a given operator $P$, there is no analytic element of
$\cV_{\cN}$ defined by Eq.~(\ref{eqn:svspc}), we can know the spectra
in the solvable sector if the anti-commutator of $\cN$-fold supercharges
constructed from $P$ can be arranged as a polynomial in the Hamiltonian
$H$ satisfying the weak quasi-solvability condition (\ref{eqn:qscon}).

\section{\label{sec:mhamBD}Mother Hamiltonians and generalized
Bender--Dunne polynomials}

\subsection{\label{ssec:puniqu}Polynomiality of Type A
Mother Hamiltonians}

Let us come back to further investigation into type A $\cN$-fold
supersymmetry.
Suppose the pair of type A $\cN$-fold supersymmetric Hamiltonians
$H_{\cN}^{\pm}=p^{2}/2+V_{\cN}^{\pm}(q)$ with the potentials
(\ref{eqn:acond1}) also satisfies the intertwining relation with
respect to an $\cM$-fold supercharge ($\cN >\cM$) given by,
\begin{align}
P_{\cM}=p^{\cM}-i\cN W(q)p^{\cM-1}+\sum_{n=0}^{\cM-2}w_{n}(q)p^{n}.
\end{align}
From a direct calculation, we have,
\begin{align}
\lefteqn{
2 i^{\cM}\left(P_{\cM}H_{\cN}^{-}-H_{\cN}^{+}P_{\cM}\right)
 =\left(\cN W''-2\cN^{2}WW'+2\cM V_{\cN}^{- '}-2w_{\cM-2}'\right)
 \partial^{\cM-1}
}\notag\\
&\equal +\sum_{k=0}^{\cM-2}\left[-2(-i)^{k-\cM}\cN W' w_{k}
 +2\binom{\cM}{k}V_{\cN}^{- (\cM-k)}+2\cN\binom{\cM-1}{k}W
  V_{\cN}^{- (\cM-1-k)}\right.\notag\\
&\equal \left.+2\sum_{n=k+1}^{\cM-2}(-i)^{n-\cM}\binom{n}{k}w_{n}
 V_{\cN}^{- (n-k)}+2(-i)^{k-1-\cM}w_{k-1}'+(-i)^{k-\cM}w_{k}''
 \right]\partial^{k}.
\end{align}
Therefore, $w_{k}(q)$ ($k=0,\dots, \cM-2$) must satisfy,
\begin{subequations}
\label{eqns:conMf}
\begin{align}
w_{\cM-2}'&=\frac{\cN}{2}W''-\cN^{2}WW'+\cM V_{\cN}^{- '},\\
w_{k-1}'&=-\sum_{n=k+1}^{\cM-2}(-i)^{n+1-k}\binom{n}{k}V_{\cN}^{- (n-k)}
 w_{n}+i\left(\frac{w_{k}''}{2}-\cN W' w_{k}\right)\notag\\
&\equal -i^{k-1-\cM}\left[\binom{\cM}{k}V_{\cN}^{- (\cM-k)}
 +\binom{\cM-1}{k}\cN WV_{\cN}^{- (\cM-1-k)}\right]\quad
 (k=1, \dots, \cM-2),
\end{align}
\end{subequations}
in addition to,
\begin{multline}
\qquad w_{0}''-2\cN W' w_{0}+2\sum_{n=1}^{\cM-2}(-i)^{n}n
 V_{\cN}^{- (n)}w_{n}\\
+2(-i)^{\cM}\left[\cM V_{\cN}^{- (\cM)}+(\cM-1)\cN W
 V_{\cN}^{- (\cM-1)}\right]=0.\qquad
\label{eqn:cnMf2}
\end{multline}
The set of the first-order differential equations (\ref{eqns:conMf})
can be integrated out in the order from $w_{\cN-2}(q)$ to $w_{0}(q)$
since all the terms appeared in the r.h.s. of Eqs.~(\ref{eqns:conMf})
becomes known functions depending only on $E(q)$ and $W(q)$ in the order.
Thus, we obtain,
\begin{align}
w_{k}(q)=f_{k}\bigl( E(q), W(q)\bigr)+C_{k}\qquad (k=0, \dots, \cM-2),
\end{align}
where $C_{k}$ are integral constants. We can put $C_{k}=0$ without
any loss of generality; if $C_{\cM'}\neq 0$ for an $\cM'$, we can always
split the $\cM$-fold supercharge as,
\begin{align}
P_{\cM}\left(C_{\cM'}\right)=P_{\cM}+C_{\cM'}P_{\cM'},
\end{align}
which means the system is also $\cM'$-fold supersymmetric.
Therefore, $w_{k}(q)$ and hence $P_{\cM}$ are uniquely determined.
On the other hand, if the type A $\cN$-fold
supersymmetric potentials (\ref{eqn:acond1}) can be
rewritten as the type A $\cM$-fold supersymmetric ones with $W(q)$
replaced by $\cN W(q)/ \cM$, that is, if the following relations,
\begin{align}
V_{\cN}^{\pm}\bigl[ E(q), W(q)\bigr]
 =V_{\cM}^{\pm}\bigl[ E(q), \cN W(q)/\cM\bigr],
\label{eqn:relNM}
\end{align}
are satisfied, it is evident that the type A $\cN$-fold supersymmetric
system is also type A $\cM$-fold supersymmetric with respect to
the following type A $\cM$-fold supercharge:
\begin{align}
P_{\cM}=\prod_{k=-(\cM-1)/2}^{(\cM-1)/2}\left(p
 -i\frac{\cN}{\cM}W(q)+ikE(q)\right).
\label{eqn:TAMsc}
\end{align}
As we have just shown, however, $P_{\cM}$ is unique and thus the
$\cM$-fold supercharge determined by the solutions of
Eqs.~(\ref{eqns:conMf}) must be the type A $\cM$-fold supercharge
Eq.~(\ref{eqn:TAMsc}). Furthermore, the constraint (\ref{eqn:cnMf2})
must be equivalent to the relation (\ref{eqn:relNM}).
Summarizing the result, we have shown that type A $\cN$-fold supercharge
is unique unless the system satisfies Eq.~(\ref{eqn:relNM}) for an $\cM$
($0<\cM <\cN$). The relation (\ref{eqn:relNM}) is equivalent to,
\begin{align}
12W(q)^{2}-\cM^{2}\left(E(q)^{2}-2E'(q)\right)
 =\frac{4}{\left(h'\right)^{2}}\left(3Q(h)^{2}+\cM^{2}H[P]\right)=0,
\end{align}
and its solutions again constitute a hyperplane $\Gamma$ in the
parameter space spanned by $\{a_{i}, b_{i}\}$. Outside the hyperplane,
the $\cN$-fold supercharge is unique and the mother Hamiltonian
is expressed as a polynomial $\mathcal{P}_{\cN}$ in the original
Hamiltonian $\bH_{\cN}$:
\begin{align}
\cH_{\cN}=\frac{1}{2}\bigl\{Q_{\cN}^{\dagger},Q_{\cN}\bigr\}
 =\mathcal{P}_{\cN}\left(\bH_{\cN}\right)\quad\text{for}
 \quad \{a_{i}, b_{i}\}\in \bbR^{5+3}\setminus\Gamma\ 
 \text{or}\ \bbC^{5+3}\setminus\Gamma.
\label{eqn:polMH}
\end{align}
On the other hand, as has been shown in Eq.~(\ref{eqn:mhplh}),
this polynomial $\mathcal{P}_{\cN}$ corresponds to
the characteristic polynomial which determines the spectra in the
solvable sector~\cite{AST2}. Since the characteristic polynomial
itself is constructed from a finite algebraic operation, it has no
discontinuity in the parameter space. Therefore, Eq.~(\ref{eqn:polMH})
must be held on $\Gamma$ and thus the type A mother Hamiltonian must
be a polynomial in the original Hamiltonian $\bH_{\cN}$ in the whole
parameter space.

\subsection{\label{ssec:relBDp}Generalized Bender--Dunne polynomials}

In 1996, Bender and Dunne introduced a set of polynomials which
determines the spectra in the solvable sector of the quasi-exactly
solvable model of case II, namely, the sextic anharmonic
oscillator \cite{BeDu1}. Soon after, Finkel et al. generalized the idea
to all the $\Lsl(2)$ quasi-exactly solvable models \cite{FiLoRo1}.
In the following, we will make a generalization without imposing
the normalizability of the solvable sector. Let the bases
$\{\phi^{\pm}\}$ of the solvable subspaces $\cV_{\cN}^{\pm}$ be,
\begin{align}
\phi_{n}^{\pm}=h^{n-1}\left(G_{\cN}U^{\mp 1}\right)^{-1}
 \qquad (n=1,\dots,\cN),
\end{align}
and the gauged Hamiltonians be,
\begin{align}
\left(G_{\cN}U^{\mp 1}\right)H_{\cN}^{\pm}\left(G_{\cN}U^{\mp 1}
 \right)^{-1}=-P(h)\frac{d^{2}}{dh^{2}}-P_{3}^{\pm}(h)
 \frac{d}{dh}-P_{2}^{\pm}(h).
\end{align}
Then we have,
\begin{align}
H_{\cN}^{\pm}\phi_{n}^{\pm}=-(n-1)(n-2)P(h)\phi_{n-2}^{\pm}
 -(n-1)P_{3}^{\pm}(h)\phi_{n-1}^{\pm}-P_{2}^{\pm}(h)\phi_{n}^{\pm}.
\label{eqn:matS2}
\end{align}
From Eqs.~(\ref{eqn:gaugH}), and (\ref{eqn:gauH+}) with (\ref{eqn:relPQ})
and (\ref{eqn:relRR}), the $P_{i}^{\pm}(h)$ are expressed as,
\begin{subequations}
\label{eqns:P23pm}
\begin{align}
P_{3}^{\pm}(h)&=-\frac{\cN-2}{2}P'(h)\mp Q(h),\\
P_{2}^{\pm}(h)&=\frac{(\cN-1)(\cN-2)}{12}P''(h)
 \pm\frac{\cN-1}{2}Q'(h)+R.
\end{align}
\end{subequations}
Substituting Eqs.~(\ref{eqn:defPh}), (\ref{eqn:defQh}) and
(\ref{eqns:P23pm}) for Eq.~(\ref{eqn:matS2}), we obtain the matrix
elements $\bS_{n,m}^{\pm}$ defined in Eq.~(\ref{eqn:matS1}):
\begin{align}
H_{\cN}^{\pm}\phi_{n}^{\pm}
&=-(n-\cN)(n-\cN+1)a_{4}\,\phi_{n+2}^{\pm}-(n-\cN)\left[
 \left(n-\frac{\cN}{2}\right)a_{3}\mp b_{2}\right]\phi_{n+1}^{\pm}
 \notag\\
&\equal -\left\{\left[(n-1)(n-\cN)+\frac{1}{6}(\cN-1)(\cN-2)\right]
 a_{2}\mp\left(n-\frac{\cN+1}{2}\right)b_{1}+R\right\}\phi_{n}^{\pm}
 \notag\\
&\equal -(n-1)\left[\left(n-\frac{\cN+2}{2}\right)a_{1}\mp b_{0}
 \right]\phi_{n-1}^{\pm}-(n-1)(n-2)a_{0}\,\phi_{n-2}^{\pm}.
\end{align}
As we have discussed in Section~\ref{sec:class}, there are five
independent type A $\cN$-fold supersymmetric models under the $GL(2,\bbC)$
transformation. By a suitable $GL(2,\bbC)$ transformation,
we can always transform $P(h)$ so that $a_{0}=0$ in all the five cases,
see Table~\ref{tab:gl2a0}. In Table~\ref{tab:gl2a0}, $e_{i}$ ($i=1,2,3$)
denote the three different single roots of the algebraic equation
$2x^{3}-g_{2}x/2-g_{3}/2=0$ and $h_{i}$ ($i=2,3$) are given by
$h_{i}=e_{i}-e_{1}$. So, we set $a_{0}=0$ hereafter. We note that case I
corresponds to the case where $P(h)$ has a quadruple root and thus
$P(h)=\text{constant}$ (quadruple root at infinity) or $P(h)\propto
(h-h_{0})^{4}$ (quadruple root at finite $h_{0}$). Therefore, in contrast
to in all the other cases, case II--V, we cannot simultaneously set
$a_{4}=0$ and $a_{0}=0$ by any $GL(2,\bbC)$ transformation in case I.
We also drop another irrelevant constant by putting $R=0$. 
\begin{table}[h]
\begin{center}
\[
\arraycolsep=5mm
\begin{array}{|c|c|c|}\hline
\text{Case} & \hat{h} & \hat{P}(h) \\
\hline
\text{I} & 1/h & h^{4}/2 \\
\hline
\text{II} & h & 2h \\
\hline
\text{III} & h & 2\nu h^{2} \\
\hline
\text{IV} & (h+2)/h & 2\nu h^{2}(h+1) \\
\hline
\text{V} & (e_{1}h+h_{2}h_{3})/h & 2h(h-h_{2})(h-h_{3}) \\
\hline
\end{array}
\]
\caption{The $GL(2,\bbC)$ transformations $\hat{h}$ which convert the
standard forms of $P(h)$ in Table~\ref{tab:class} into other forms
$\hat{P}(h)$ satisfying $a_{0}=0$.}
\label{tab:gl2a0}
\end{center}
\end{table}

In the next, we introduce a set of functions $P_{n}^{[\cN]}(E)$
by putting a solution of the Schr\"{o}dinger equation as follows:
\begin{align}
\psi^{\pm}=\left\{
\begin{array}{ll}
\displaystyle{\sum_{n=0}^{\infty}\frac{P_{n}^{[\cN]}(E)}{
 (\pm b_{0})^{n}\,n!}\,\phi_{n+1}^{\pm}} & \text{if}\quad a_{1}=0,\\
\displaystyle{\sum_{n=0}^{\infty}\frac{P_{n}^{[\cN]}(E)}{
 (-a_{1})^{n}\,n!\,\Gamma\left(n-\frac{\cN-2}{2}\mp\frac{b_{0}}{a_{1}}
 \right)}\,\phi_{n+1}^{\pm}}\quad & \text{if}\quad a_{1}\neq 0.
\end{array}
\right.
\end{align}
From the requirement that the above $\psi^{\pm}$ satisfies $H_{\cN}^{\pm}
\psi^{\pm}=E\psi^{\pm}$, we obtain a four-term recursion
relation for $P_{n}^{[\cN]}(E)$:
\begin{align}
P_{n+1}^{[\cN]}(E)
&=\bigl(E+A_{n}^{[\cN]}\bigr)P_{n}^{[\cN]}(E)
 -n(n-\cN)B_{n}^{[\cN]}P_{n-1}^{[\cN]}(E)\notag\\
&\equal +n(n-1)(n-\cN)(n-\cN-1)C_{n}^{[\cN]}P_{n-2}^{[\cN]}(E),
\label{eqn:4recP}
\end{align}
where $A_{n}^{[\cN]}$, $B_{n}^{[\cN]}$ and $C_{n}^{[\cN]}$ are given by,
\begin{subequations}
\label{eqns:dfABC}
\begin{align}
A_{n}^{[\cN]}&=\left[n(n-\cN+1)+\frac{1}{6}(\cN-1)(\cN-2)\right]a_{2}
 \mp\left(n-\frac{\cN-1}{2}\right)b_{1},\\
B_{n}^{[\cN]}&=\left[\left(n-\frac{\cN}{2}\right)a_{1}\mp b_{0}\right]
 \left[\left(n-\frac{\cN}{2}\right)a_{3}\mp b_{2}\right],\\
C_{n}^{[\cN]}&=a_{4}\left[\left(n-\frac{\cN}{2}\right)a_{1}\mp b_{0}\right]
 \left[\left(n-\frac{\cN+2}{2}\right)a_{1}\mp b_{0}\right].
\end{align}
\end{subequations}
We can set $P_{0}^{[\cN]}(E)=1$ without any loss of generality.
Then, each $P_{n}^{[\cN]}(E)$ generated by Eq.~(\ref{eqn:4recP})
becomes a monic polynomial of $n$th
degree. We call it a generalized Bender--Dunne polynomial (GBDP).
In the case of $a_{4}=0$, Eq.~(\ref{eqn:4recP}) reduces to a three-term
recursion relation with (in general) $n(n-\cN)B_{n}^{[\cN]}\neq 0$
for $n\neq\cN$ and thus the set of $P_{n}^{[\cN]}(E)$ forms a
\textit{weakly} orthogonal polynomial system~\cite{Chiha}. Therefore,
case I is special in the sense that the set of polynomials
$P_{n}^{[\cN]}$ does not form a weakly orthogonal polynomial system.
One of the most notable properties of the GBDP is the factorization
property; due to the structure of Eq.~(\ref{eqn:4recP}), all the
polynomials $P_{n}^{[\cN]}(E)$ for $n\ge\cN$ are factorized as,
\begin{align}
P_{\cN+n}^{[\cN]}(E)=Q_{n}^{[\cN]}(E)P_{\cN}^{[\cN]}(E)
 \qquad (n\ge 0),
\label{eqn:facto}
\end{align}
where $Q_{n}^{[\cN]}(E)$ is a polynomial of degree $n$ satisfying
another four-term recursion relation:
\begin{align}
Q_{n+1}^{[\cN]}(E)
&=\bigl(E+A_{\cN+n}^{[\cN]}\bigr)Q_{n}^{[\cN]}(E)
 -n(n+\cN)B_{\cN+n}^{[\cN]}Q_{n-1}^{[\cN]}(E)\notag\\
&\equal +n(n-1)(n+\cN)(n+\cN-1)C_{\cN+n}^{[\cN]}Q_{n-2}^{[\cN]}(E).
\label{eqn:4recQ}
\end{align}
In the case of $a_{4}=0$, Eq.~(\ref{eqn:4recQ}) reduces to a three-term
recursion relation with (in general) $n(n+\cN)B_{\cN+n}^{[\cN]}\neq 0$
for all $n>0$ and thus the set of $Q_{n}^{[\cN]}(E)$ forms an orthogonal
polynomial system~\cite{Chiha}. Therefore, case I is again special in
the sense that the set of polynomials $Q_{n}^{[\cN]}$ does not form
an orthogonal polynomial system. Due to the factorization property
(\ref{eqn:facto}), $P_{\cN}^{[\cN]}(E)$ reserves special status
among the GBDPs. We call $P_{\cN}^{[\cN]}(E)$ $\cN$-th
\textit{critical} GBDP after the terminology in Ref.~\cite{BeDu1}.

When $E$ takes one of the spectral values $E_{n}$ ($n=1,\dots,\cN$)
in the solvable sector, $\psi^{\pm}$ must be an element of
$\cV_{\cN}^{\pm}$.
From the factorization property (\ref{eqn:facto}), the condition
$\psi^{\pm}\in\cV_{\cN}^{\pm}$ is fulfilled if and only if,
\begin{align}
P_{\cN}^{[\cN]}(E_{n})=0\qquad (n=1,\dots,\cN).
\end{align}
This means that all the zeros of the $\cN$-th critical GBDP
must correspond to the spectra in the solvable sector of the
quasi-solvable Hamiltonians $H_{\cN}^{\pm}$. On the other hand,
they are also given by solutions of the characteristic equation
(\ref{eqn:chaeq}).
Therefore, each of the critical GBDP is proportional to the corresponding
characteristic polynomial. Comparing the coefficients of the
highest-degree term with each other, we have,
\begin{align}
\det\bM_{\cN}^{\pm}(E)=2^{\cN}{P}_{\cN}^{[\cN]}(E).
\label{eqn:detBD}
\end{align}
As we have proved before, the type A mother Hamiltonian is expressed
solely by the characteristic polynomial of the original Hamiltonian as
Eq.~(\ref{eqn:mhplh}). Combining Eq.~(\ref{eqn:mhplh}) with
(\ref{eqn:detBD}), we finally obtain an intriguing relation:
\begin{align}
\cH_{\cN}=\frac{1}{2}\det\bM_{\cN}^{\pm}(\bH_{\cN})
 =2^{\cN-1}P_{\cN}^{[\cN]}(\bH_{\cN}).
\label{eqn:MHaBD}
\end{align}
Furthermore, remembering that the mother Hamiltonian is defined
by the anti-commutator of the $\cN$-fold supercharges (\ref{eqn:moham}),
we get the complete type A $\cN$-fold superalgebra:
\begin{subequations}
\begin{gather}
\bigl[ Q_{\cN}, \bH_{\cN}\bigr]
 =\bigl[ Q_{\cN}^{\dagger}, \bH_{\cN}\bigr]=0,\\
\bigl\{ Q_{\cN}, Q_{\cN}\bigr\}
 =\bigl\{ Q_{\cN}^{\dagger}, Q_{\cN}^{\dagger}\bigr\}=0,\\
\bigl\{ Q_{\cN}^{\dagger}, Q_{\cN}\bigr\}
 =2^{\cN}P_{\cN}^{[\cN]}(\bH_{\cN}).
\end{gather}
\end{subequations}

\subsection{\label{ssec:exmham}Examples}

In order to confirm the previous argument,
we will show the explicit results for $\cN=1,2,3$.
By solving the recursion relation (\ref{eqn:4recP}), we obtain
$P_{n}^{[\cN]}(E)$ ($n\le\cN$) as follows:\\

1) $\cN=1$:
\begin{align}
P_{1}^{[1]}(E)=E,
\label{eqn:crit1}
\end{align}

2) $\cN=2$:
\begin{subequations}
\begin{align}
P_{1}^{[2]}(E)&=E\pm\frac{b_{1}}{2},\\
P_{2}^{[2]}(E)&=E^{2}+b_{0}b_{2}-\frac{b_{1}^{2}}{4},
\label{eqn:crit2}
\end{align}
\end{subequations}

3) $\cN=3$:
\begin{subequations}
\begin{align}
P_{1}^{[3]}(E)&=E+\frac{a_{2}}{3}\pm b_{1},\\
P_{2}^{[3]}(E)&=E^{2}-\frac{1}{3}\left(a_{2}\mp 3b_{1}\right)E\notag\\
&\equal -\frac{1}{18}\left(4a_{2}^{2}-9a_{1}a_{3}\mp 18a_{1}b_{2}
 \pm 12a_{2}b_{1}\mp 18a_{3}b_{0}-36b_{0}b_{2}\right),\\
P_{3}^{[3]}(E)&=E^{3}+\frac{1}{3}\left(3a_{1}a_{3}-a_{2}^{2}
 +12b_{0}b_{2}-3b_{1}^{2}\right)E
 -\frac{1}{27}\left(2a_{2}^{3}-9a_{1}a_{2}a_{3}\right.\notag\\
&\equal \left.+27a_{1}^{2}a_{4}-108a_{4}b_{0}^{2}+54a_{3}b_{0}b_{1}
 -18a_{2}b_{1}^{2}-36a_{2}b_{0}b_{2}+54a_{1}b_{1}b_{2}\right).
\label{eqn:crit3}
\end{align}
\end{subequations}
On the other hand, the direct calculation of the mother Hamiltonians
reads as follows:\\

1) $\cN=1$:
\begin{align}
2\cH_{1}=2\bH_{1},
\label{eqn:mham1}
\end{align}

2) $\cN=2$:
\begin{align}
2\cH_{2}=4\left(\bH_{2}\right)^{2}+D_{2}[Q],
\label{eqn:mham2}
\end{align}

3) $\cN=3$:
\begin{align}
2\cH_{3}=8\left(\bH_{3}\right)^{3}-\frac{8}{3}\bigl(i_{2}[P]-3D_{2}[Q]
 \bigr)\bH_{3}+\frac{16}{27}\bigl(j_{3}[P]+9I_{1,2}[P,Q]\bigr).
\label{eqn:mham3}
\end{align}
In the above, $D_{2}$, $i_{2}$, $j_{3}$ and $I_{1,2}$ are the absolute
invariants \textit{expressed in terms of} $E(q)$ \textit{and} $W(q)$,
see Eqs.~(\ref{eqn:D2})--(\ref{eqn:I12}).
Comparing the critical GBDPs, Eqs.~(\ref{eqn:crit1}), (\ref{eqn:crit2}),
and (\ref{eqn:crit3}), obtained by solving the recursion relation
(\ref{eqn:4recP}) with the mother Hamiltonians (\ref{eqn:mham1})--%
(\ref{eqn:mham3}), and noting that we have
put $a_{0}=0$, we confirm the relation (\ref{eqn:MHaBD})
for $\cN=1,2,3$. We also note that since the mother Hamiltonians are
invariant under the $GL(2,K)$ transformation, all the coefficients of
the critical GBDPs should be expressed solely in terms of
the absolute invariants listed in Eq.~(\ref{eqn:abinv}), as the above
examples indicate. From the facts that all the GBDPs are homogeneous
polynomials in $E$, $a_{i}$ and $b_{i}$ due to the structure of the
recursion relation (\ref{eqn:4recP}), and that all the critical GBDPs
are symmetric under the transformation $b_{i}\rightarrow -b_{i}$,
the general form of the critical GBDPs should be,
\begin{align}
P_{\cN}^{[\cN]}(E)&=E^{\cN}+\sum_{k=0}^{\cN-2}E^{k}\Biggl[\sum_{n_{1},
 \dots, n_{6}}\delta_{2n_{1}+2n_{2}+3n_{3}+3n_{4}+4n_{5}+12n_{6},\,\cN-k}
 \times\notag\\
&\equal \times C_{k;\, n_{1},\dots, n_{6}}^{[\cN]}\left(D_{2}\right)^{n_{1}}
 \left(i_{2}\right)^{n_{2}}\left(j_{3}\right)^{n_{3}}\left(I_{1, 2}
 \right)^{n_{4}}\left(I_{2, 2}\right)^{n_{5}}\left(I_{3, 3}\right)^{2n_{6}}
 \Biggr].
\label{eqn:cgbdp}
\end{align}
where $C_{k;\, n_{1},\dots, n_{6}}^{[\cN]}$ are constants.
For example, the first five are,
\begin{subequations}
\begin{align}
P_{1}^{[1]}(E)&=E,\\
P_{2}^{[2]}(E)&=E^{2}+\frac{1}{4}D_{2},\\
P_{3}^{[3]}(E)&=E^{3}-\frac{1}{3}\left(i_{2}-3D_{2}\right)E
 +\frac{2}{27}\left(j_{3}+9I_{1, 2}\right),\\
P_{4}^{[4]}(E)&=E^{4}-\frac{1}{2}\left(4i_{2}-5D_{2}\right)E^{2}
 +4I_{1, 2}\, E+\left(i_{2}\right)^{2}+2I_{2, 2}+i_{2}D_{2}+\frac{9}{16}
 \left(D_{2}\right)^{2},\\
P_{5}^{[5]}(E)&=E^{5}-\left(7i_{2}-5D_{2}\right)E^{3}-2\left(j_{3}
 -7I_{1, 2}\right)E^{2}\notag\\
&\equal +4\left(3\left(i_{2}\right)^{2}+4I_{2, 2}\right)E
 +8\left(i_{2}-D_{2}\right)\left(j_{3}+I_{1, 2}\right).
\end{align}
\end{subequations}
Since the zeros of the critical GBDPs correspond to the spectra
of $H_{\cN}^{\pm}$ in the solvable sector, these spectra can be regarded
as functions of the six absolute invariants. Furthermore, as an
interpretation of Eq.~(\ref{eqn:cgbdp}), the critical GBDPs, and
equivalently the type A mother Hamiltonians, can be
regarded as generating functions of the absolute invariants composed
of $P(h)$ and $Q(h)$.

\section{\label{sec:concl}Concluding Remarks}

In this article, we have fully investigated general aspects of type A
$\cN$-fold supersymmetry. The two different approaches in
Section~\ref{sec:typea} reveal both the analytic and algebraic
structures of the systems. The intimate relation between the algebraic
forms of the pair Hamiltonians, Eqs.~(\ref{eqn:gsolH}) and
(\ref{eqn:psolH}), provides an interesting problem. Suppose we have
a quasi-solvable system $\tH$ which is represented by a quadratic
polynomial of a set of first-order differential operators constituting
a finite dimensional representation of a Lie algebra. If we construct
another system $\bar{H}$ from $\tH$ by interchanging the order of the
quadratic terms and interchanging the sign of the coefficients of the
linear terms, do $\tH$ and $\bar{H}$ (with suitable gauge transformations)
always form an $\cN$-fold supersymmetric pair? This problem may be
extended to quasi-solvable many-body systems~\cite{Tanak1,Tanak2}.

The invariance of the Hamiltonians as well as the type A $\cN$-fold
supercharge under the $GL(2,K)$ transformation enable us to obtain
the complete classification of the type A models. This invariance also
plays an essential role in generalizing the Bender--Dunne polynomials to
for all the type A models. With regard to the issue, we should stress
that, up to now, we have found no reason that a set of the polynomials
should be weakly orthogonal, although all the existing papers concerning
about the issue~\cite{BeDu1,KrUsWa1,FiLoRo1,KhMa1,KhMa2,FiLoRo2,Gangu2},
as far as we know,
have required weak orthogonality by imposing, for instance, the
normalizabiliy, or restricted the considerations in which weak
orthogonality is fulfilled. One of the most essential properties that
sets of polynomials associated with quasi-solvable systems should
share is the factorization property (\ref{eqn:facto}). The most general
form of a recursion relation for a set of polynomials $P_{n}^{[\cN]}$
which guarantee the factorization property may be the following:
\begin{align}
P_{n+1}^{[\cN]}(E)=\bigl(E+A_{0,n}^{[\cN]}\bigr)P_{n}^{[\cN]}(E)
+\sum_{k=1}^{K}\left[(-)^{k}\prod_{l=0}^{k-1}(n-l)(n-\cN-l)
 A_{k,n}^{[\cN]}P_{n-k}^{[\cN]}(E)\right].
\end{align}
Then, an interesting question is, what conditions a set of the
coefficients $\{ A_{k,n}^{[\cN]}\}$ should satisfy for the existence
of a quasi-solvable system whose spectra in its solvable sector are
given by the zeros of the critical polynomial $P_{\cN}^{[\cN]}(E)$
obtained by the above recursion relation. This question may provide
an alternative way to find out a new quasi-solvable and also a new
$\cN$-fold supersymmetric model.

\begin{acknowledgments}
The author would like to thank H.~Aoyama, F.~Finkel,
D.~G\'{o}mez-Ullate, A.~Gonz\'{a}lez-L\'{o}pez, N.~Nakayama,
C.~Quesne, M.~A.~Rodr\'{\i}guez, R.~Sasaki, M.~Sato and K.~Takasaki
for useful discussions.
The author would also like to thank all the members of Departamento
de F\'{\i}sica Te\'{o}rica II, Universidad Complutense, for their
kind hospitality during his stay.
This work was supported in part by a JSPS research fellowship.
\end{acknowledgments}

\appendix

\section{\label{app:invsp}Invariants of a System of Polynomials}

Since the type A $\cN$-fold supersymmetric models have the underlying
$GL(2,K)$ invariance discussed in Section~\ref{sec:class}, all the
relevant quantities should be expressed in terms of the covariants
and invariants under the transformation. In this appendix, we summarize
the covariants and invariants of a system of polynomials needed in
this article.
For more details, see Ref.~\cite{LoKaOl3} and references cited therein.

The multiplier representation $\rho_{m,\, i}$ of $GL(2,K)$ on the space
of polynomials of degree at most $m$ is defined by,
\begin{align}
F(h)\mapsto \hat{F}(h)=\Delta^{i}(\gamma h+\delta)^{m}F(\hat{h}),
\label{eqn:mlrep}
\end{align}
where $\hat{h}$ and $\Delta$ are defined by Eq.~(\ref{eqn:lftoh}).
For example, we see from Eqs.~(\ref{eqns:tfPaQ}) that $P(h)$ belongs
to the representation $\rho_{4,\, -2}$ while $Q(h)$ belongs to the
representation $\rho_{\, 2,\, -1}$.

In order to construct a complete system of covariants and invariants,
the process of transvection is essential. Let $F(h)$ be a polynomial
of degree at most $m$ belonging to the representation $\rho_{m,\, i}$
and $G(h)$ be a polynomial of degree at most $n$ belonging to the
representation $\rho_{n,\, j}$. The $r$th transvectant of $F$ and $G$
is defined by,
\begin{align}
(F,G)^{(r)}=\sum_{k=0}^{r}(-1)^{k}\binom{r}{k}
 \frac{(m-r+k)!(n-k)!}{(m-r)!(n-r)!}F^{(r-k)}G^{(k)}.
\end{align}
Then, $(F,G)^{(r)}$ is a polynomial of degree at most $m+n-2r$
belonging to the representation $\rho_{m+n-2r,\, i+j+r}$.

For a single quadratic polynomial $Q$, there are two independent
covariants, namely, $Q$ itself and its discriminant $D_{2}$ given by,
\begin{align}
D_{2}[Q]=\frac{1}{2}(Q,Q)^{(2)}.
\end{align}
For a single quartic polynomial $P$, there are five independent
covariants, namely, $P$ itself, the Hessian $H$ and the Jacobian
$J$ given by,
\begin{subequations}
\begin{align}
H[P]&=\frac{1}{24}(P,P)^{(2)}=PP''-\frac{3}{4}\left(P'\right)^{2},\\
J[P]&=\frac{1}{24}(P,H)^{(1)}=-\frac{1}{24}\left[
 4P^{2}P^{(3)}-6PP'P''+3\left(P'\right)^{3}\right],
\end{align}
\end{subequations}
and the two invariants given by,
\begin{align}
i_{2}[P]=\frac{1}{96}(P,P)^{(4)},\qquad j_{3}[P]=\frac{1}{96}(P,H)^{(4)}.
\end{align}
The discriminant of $P$ is expressed in terms of the above invariants
as $D_{6}[P]=i_{2}[P]^{3}-j_{3}[P]^{2}$.

According to the invariant theory of polynomials, the complete list
of independent covariants for the system consisting of a quartic
and a quadratic polynomial is the following:\\

1) Absolute invariants $\rho_{\, 0,\, 0}$:
\begin{align}
D_{2}[Q],\quad i_{2}[P],\quad j_{3}[P],\quad I_{1, 2}[P,Q],\quad
 I_{2, 2}[P,Q],\quad I_{3, 3}[P,Q],
\label{eqn:abinv}
\end{align}

2) Quadratic covariants $\rho_{\, 2,\, -1}$:
\begin{align}
Q,\quad (P,Q)^{(2)},\quad (H,Q)^{(2)},\quad
 (P,Q^{2})^{(3)},\quad (H,Q^{2})^{(3)},\quad (J,Q^{2})^{(4)},
\label{eqn:2cova}
\end{align}

3) Quartic covariants $\rho_{4,\, -2}$:
\begin{align}
P,\quad H[P],\quad (P,Q)^{(1)},\quad (H,Q)^{(1)},\quad (J,Q)^{(2)},
\label{eqn:4cova}
\end{align}

4) Sextic covariant $\rho_{\, 6,\, -3}$:
\begin{align}
J[P].
\label{eqn:6cova}
\end{align}
In the above, the absolute invariants $I_{1, 2}$, $I_{2, 2}$ and
$I_{3, 3}$ are given by,
\begin{align}
I_{1, 2}[P,Q]=\frac{1}{96}(P,Q^{2})^{(4)},\quad I_{2, 2}[P,Q]=\frac{1}{96}
 (H,Q^{2})^{(4)},\quad I_{3, 3}[P,Q]=\frac{1}{64800}(J,Q^{3})^{(6)}.
\end{align}
We can express the above quantities in terms of $h(q)$, $E(q)$ and $W(q)$
with the aid of the following relations derived from
Eqs.~(\ref{eqn:defEq}) and (\ref{eqns:cano-}):
\begin{subequations}
\begin{gather}
P=\frac{1}{2}\left(h'\right)^{2},\quad P'=h''=Eh',
 \quad P''=E'+E^{2},\\
P^{(3)}=\frac{1}{h'}\left(E''+2EE'\right),
 \quad P^{(4)}=\frac{1}{\left(h'\right)^{2}}
 \left[E^{(3)}+EE''+2\left(E'\right)^{2}-2E^{2}E'\right],
\end{gather}
\end{subequations}
\begin{subequations}
\begin{gather}
Q=-Wh',\qquad Q'=-\left(W'+EW\right),\\
Q''=-\frac{1}{h'}\left(W''+EW'+E'W\right).
\end{gather}
\end{subequations}
For example, the Hessian $H$ and the Jacobian $J$ are expressed as,
\begin{align}
H[P]=\frac{\left(h'\right)^{2}}{4}\left(2E'-E^{2}\right),
 \qquad J[P]=-\frac{\left(h'\right)^{3}}{24}\left(E''-EE'\right).
\end{align}
Among the covariants (\ref{eqn:abinv})--(\ref{eqn:6cova}), the absolute
invariants play an essential role in Section~\ref{sec:mhamBD}. In the
following, we show the explicit forms of them in terms of the
coefficients of the polynomials $P(h)$ and $Q(h)$, as well as in terms
of $E(q)$ and $W(q)$ (for the first four invariants because the
expressions for the last two are lengthy):
\begin{align}
D_{2}[Q]&=2QQ''-\left(Q'\right)^{2}\notag\\
&=2WW''-\left(W'\right)^{2}+\left(2E'-E^{2}\right)W^{2}\notag\\
&=4b_{0}b_{2}-b_{1}^{2},
\label{eqn:D2}
\end{align}
\begin{align}
i_{2}[P]&=\frac{1}{4}\left[2PP^{(4)}
 -2P'P^{(3)}+\left(P''\right)^{2}\right]\notag\\
&=\frac{1}{4}\left[E^{(3)}-EE''+3\left(E'\right)^{2}
 -4E^{2}E'+E^{4}\right]\notag\\
&=12a_{0}a_{4}-3a_{1}a_{3}+a_{2}^{2},
\end{align}
\begin{align}
2j_{3}[P]&=\frac{1}{8}\left[ 12PP''P^{(4)}
 -6P\left(P^{(3)}\right)^{2}-9\left(P'\right)^{2}P^{(4)}
 +6P'P''P^{(3)}-2\left(P''\right)^{3}\right]\notag\\
&=\frac{1}{8}\left[6E'E^{(3)}-3\left(E''\right)^{2}-3E^{2}E^{(3)}
 \right.\notag\\
&\equal\left. +10\left(E'\right)^{3}+3E^{3}E''
 -24E^{2}\left(E'\right)^{2}+12E^{4}E'-2E^{6}\right]\notag\\
&=72a_{0}a_{2}a_{4}-27a_{0}a_{3}^{2}-27a_{1}^{2}a_{4}+9a_{1}a_{2}a_{3}
 -2a_{2}^{3},
\end{align}
\begin{align}
4I_{1, 2}[P,Q]
&=P^{(4)}Q^{2}-2P^{(3)}QQ'+2P''\left[QQ''+\left(Q'\right)^{2}\right]
 -6P'Q'Q''+6P\left(Q''\right)^{2}\notag\\
&=\left[E^{(3)}-EE''+7\left(E'\right)^{2}-8E^{2}E'+2E^{4}\right] W^{2}
 \notag\\
&\equal -2\left(E''-EE'\right)WW'+\left(2E'-E^{2}\right)\left[
 4WW''+\left(W'\right)^{2}\right]+3\left(W''\right)^{2}\notag\\
&=4\left(6a_{4}b_{0}^{2}-3a_{3}b_{0}b_{1}+2a_{2}b_{0}b_{2}+a_{2}b_{1}^{2}
 -3a_{1}b_{1}b_{2}+6a_{0}b_{2}^{2}\right),
\label{eqn:I12}
\end{align}
\begin{align}
2I_{2, 2}[P,Q]
&=\frac{1}{2}\left\{H^{(4)}Q^{2}-2H^{(3)}QQ'+2H''\left[QQ''+\left(Q'
 \right)^{2}\right]-6H'Q'Q''+6H\left(Q''\right)^{2}\right\}\notag\\
&=3\left(8a_{2}a_{4}-3a_{3}^{2}\right)b_{0}^{2}-6\left(6a_{1}a_{4}
 -a_{2}a_{3}\right)b_{0}b_{1}\notag\\
&\equal +\left(24a_{0}a_{4}+3a_{1}a_{3}-2a_{2}^{2}\right)
 \left(2b_{0}b_{2}+b_{1}^{2}\right)-6\left(6a_{0}a_{3}-a_{1}a_{2}\right)
 b_{1}b_{2}\notag\\
&\equal +3\left(8a_{0}a_{2}-3a_{1}^{2}\right)b_{2}^{2},
\end{align}
\begin{align}
I_{3, 3}[P,Q]
&=\frac{1}{90}\left\{J^{(6)}Q^{3}-3J^{(5)}Q^{2}Q'+3J^{(4)}Q
 \left[QQ''+2\left(Q'\right)^{2}\right]-6J^{(3)}Q'\left[3QQ''+\left(
 Q'\right)^{2}\right]\right.\notag\\
&\equal \left.+18J''Q''\left[QQ''+2\left(Q'\right)^{2}\right]
 -90J'Q'\left(Q''\right)^{2}+90J\left(Q''\right)^{3}\right\}\notag\\
&=\left(8a_{1}a_{4}^{2}-4a_{2}a_{3}a_{4}+a_{3}^{3}\right)b_{0}^{3}
 -\left(16a_{0}a_{4}^{2}+2a_{1}a_{3}a_{4}-4a_{2}^{2}a_{4}+a_{2}a_{3}^{2}
 \right)b_{0}^{2}b_{1}\notag\\
&\equal +\left(8a_{0}a_{3}a_{4}-4a_{1}a_{2}a_{4}+a_{1}a_{3}^{2}\right)
 \left(b_{0}^{2}b_{2}+b_{0}b_{1}^{2}\right)-6\left(a_{0}a_{3}^{2}
 -a_{1}^{2}a_{4}\right)b_{0}b_{1}b_{2}\notag\\
&\equal -\left(a_{0}a_{3}^{2}-a_{1}^{2}a_{4}\right)b_{1}^{3}
 -\left(8a_{0}a_{1}a_{4}-4a_{0}a_{2}a_{3}+a_{1}^{2}a_{3}\right)
 \left(b_{0}b_{2}^{2}+b_{1}^{2}b_{2}\right)\notag\\
&\equal +\left(16a_{0}^{2}a_{4}+2a_{0}a_{1}a_{3}-4a_{0}a_{2}^{2}
 +a_{1}^{2}a_{2}\right)b_{1}b_{2}^{2}-\left(8a_{0}^{2}a_{3}
 -4a_{0}a_{1}a_{2}+a_{1}^{3}\right)b_{2}^{3}.
\end{align}


\end{document}